\documentclass[useAMS,usenatbib]{mn2e}

\usepackage{graphicx}
\usepackage{epsfig}

\title[The 2SLAQ Luminous Red Galaxy Survey]
      {The 2dF-SDSS LRG and QSO (2SLAQ) Luminous Red Galaxy Survey}

\author[Russell Cannon et al.]  {Russell Cannon$^{1}$\thanks{E-mail:
	rdc@aao.gov.au}, Michael Drinkwater$^{2}$, Alastair Edge$^3$,
	Daniel Eisenstein$^4$, \newauthor Robert Nichol$^5$, Phillip
	Outram$^3$, Kevin Pimbblet$^2$, Roberto De Propris$^{6}$,
	\newauthor Isaac Roseboom$^2$, David Wake$^{3,5}$, Paul
	Allen$^7$, Joss Bland-Hawthorn$^1$, \newauthor Terry
	Bridges$^8$, Daniel Carson$^5$, Kuenley Chiu$^9$, Matthew
	Colless$^1$, \newauthor Warrick Couch$^{10}$, Scott Croom$^1$,
	Simon Driver$^6$, Stephen Fine$^{11}$, Paul Hewett$^{12}$,
	\newauthor Jon Loveday$^{13}$, Nicholas Ross$^3$, Elaine
	M. Sadler$^{11}$, Tom Shanks$^3$, Robert Sharp$^1$, \newauthor
	J. Allyn Smith$^{14}$, Chris Stoughton$^{15}$, Peter
	Weilbacher$^{3,16}$, Robert J. Brunner$^{17}$, \newauthor
	Avery Meiksin$^{18}$, Donald P. Schneider$^{19}$\\
$^{1}$ Anglo-Australian Observatory, PO Box 296, Epping, NSW 1710,
Australia\\
$^{2}$ Department of Physics, University of Queensland, QLD 4072, 
Australia\\
$^{3}$ Department of Physics, University of Durham,
South Road, Durham DH1 3LE, UK\\
$^{4}$ Steward Observatory, 933 N. Cherry Ave, Tucson, AZ 85721, USA\\
$^{5}$ Institute of Cosmology and Gravitation, University of Portsmouth,
Portsmouth, PO1 2EG, UK\\
$^{6}$ Cerro Tololo Inter-American Observatory, Casilla 63-D, La Serena, 
Chile\\
$^{7}$ Research School of Astronomy and Astrophysics, Australian 
National University, Canberra, ACT 2600, Australia\\
$^{8}$ Physics Department, Queen's University, Kingston, Ontario, Canada
K7M 3N6\\
$^{9}$ Dept. of Physics \& Astronomy, The Johns Hopkins University,
3400 North Charles Street, Baltimore, MD 21218, USA\\
$^{10}$ School of Physics, University of New South Wales, Sydney, NSW 2052,
Australia\\
$^{11}$ School of Physics, University of Sydney, NSW 2006, Australia\\ 
$^{12}$ Institute of Astronomy, Madingley Road, Cambridge CB3 0HA, UK\\
$^{13}$ Astronomy Centre, University of Sussex, Falmer, Brighton BN1 9QJ, 
UK\\
$^{14}$ Los Alamos National Laboratory, ISR-4, MS D448, Los Alamos, NM
87544-1724.;\\    Department of Physics and Astronomy, University of
Wyoming, P.O. Box 3905, Laramie, WY 82071, USA\\
$^{15}$ Fermi National Accelerator Laboratory, PO Box 500, Batavia, IL
60510, USA\\
$^{16}$ Astrophysikalisches Institut Potsdam, An der Sternwarte 16,
D-14482, Potsdam, Germany\\
$^{17}$ Department of Astronomy, University of Illinois, 1002 W. Green St.,
Urbana, IL 61801, USA\\
$^{18}$ Department of Astronomy, University of Edinburgh, Blackford
Hill, Edinburgh EH9 3HJ, UK\\
$^{19}$ Department of Astronomy, The Pennsylvania State University,
University Park, PA 16802, USA}
\begin{document}

\date{Accepted ... . Received 2006 July 25; in original form 2006
February 13}

\pagerange{\pageref{firstpage}--\pageref{lastpage}} \pubyear{2006}

\maketitle

\label{firstpage}

\begin{abstract}

We present a spectroscopic survey of almost 15,000 candidate
intermediate-redshift Luminous Red Galaxies (LRGs) brighter than $i =
19.8$, observed with 2dF on the Anglo-Australian Telescope.  The
targets were selected photometrically from the Sloan Digital Sky
Survey (SDSS) and lie along two narrow equatorial strips covering
180~$\rm{deg}^2$.  Reliable redshifts were obtained for 92\% of the
targets and the selection is very efficient: over 90\% have
$0.45<z<0.8$.  More than 80\% of the $\sim 11,000$ red galaxies have
pure absorption-line spectra consistent with a passively-evolving old
stellar population.  The redshift, photometric and spatial
distributions of the LRGs are described.  The 2SLAQ data will be
released publicly from mid-2006, providing a powerful resource for
observational cosmology and the study of galaxy evolution.

\end{abstract}


\section{Introduction}

The Two-degree Field Galaxy Redshift Survey (2dFGRS: \citealt{cdm01})
and the MAIN galaxy sample \citep{str02} of the Sloan Digital Sky
Survey (SDSS: \citealt{yor00}), have provided detailed maps of the
local structure from hundreds of thousands of galaxies of all types at
a mean redshift of $z \sim 0.1$, while the SDSS is also generating a
large sample of more distant Luminous Red Galaxies (LRGs) with
$0.25<z<0.5$ \citep{eis01}.  The present paper describes a new survey
which combines the precision of the SDSS photometric survey with the
spectroscopic capabilities of the Two-degree Field instrument (2dF) on
the 3.9m Anglo-Australian Telescope (AAT), to extend the LRG survey
from $z \approx 0.45$ to $z \approx 0.7$.  LRGs constitute ideal
tracers of large-scale structure at intermediate redshifts ($0.3 < z <
1$) because they are intrinsically luminous and spectroscopically
homogeneous, and can be reliably identified photometrically.  LRGs are
strongly clustered, with a bias towards high density regions in ways
that are believed to be well understood.

The shift in the focus of observational cosmology from determining
cosmological parameters to attempting to constrain galaxy evolution
and formation has accelerated over the past decade, especially since
the release of the first WMAP results \citep{sper03}.  The
spectroscopic information extracted from the 2dFGRS and SDSS has
provided important clues as to the nature of galaxies in the local
Universe and their recent history \citep{kau03, ah04}.  Although
galaxies are generally believed to form hierarchically through a
succession of mergers, it appears that the many of the most massive
elliptical galaxies are very old as judged from their stellar content.
Thus it seems that the rate of the merger process depends strongly on
the local density.  A large survey of LRGs at $z \sim 0.55$ therefore
has two objectives: revealing the large scale structure and clustering
of matter when the universe was about two thirds of its present age,
and understanding the evolution of the most massive galaxies
themselves.

The SDSS LRG Survey \citep{eis01} is limited to $r<19.5$ by the fixed
exposure time of 45m, set for SDSS spectroscopic observations in the
lower redshift MAIN galaxy survey.  This is well above the limit to
which LRGs can be selected reliably from the SDSS imaging data.  The
original 2dFGRS included all galaxies with blue magnitude $b_{\rm J} <
19.5$, which reached $z=0.3$ and required exposures of less than an
hour on the AAT.  By increasing the exposure time to 4h and targetting
only the relatively rare LRGs for spectroscopy in the red spectral
region, 2dF can reach $i=19.8$ and $z\sim0.7$.  

The surface density of LRG targets requires only 200 of the 400 2dF
fibres within each $2\degr$ diameter field, so all the LRGs were
observed in one of the two 2dF spectrographs.  The other was used for
a lower resolution survey of faint QSOs \citep{scr06, ric05} which
extends the earlier 2dF QSO Redshift survey (2QZ: \citealt{scr04}).
The two new surveys together comprise the 2dF-SDSS LRG And QSO survey,
2SLAQ.  A bonus of 2SLAQ is that there is some overlap in redshift
range for the LRGs and QSOs, which will enable a direct comparison
between their spatial distributions.  There is also scope for
investigating environmental and gravitational lensing effects, through
the comparison of foreground LRGs and the absorption spectra of nearby
background QSOs (cf. \citealt{b04}).

The 2SLAQ targets were selected within two narrow equatorial strips,
each $2\degr$ wide and about $100\degr$ long running through the
northern and southern Galactic caps, chosen for having good photometry
at the time of the first SDSS data release (DR1: \citealt{ab03}).  The
aim of the LRG survey was to obtain spectra for 10,000 galaxies in the
redshift range $0.45<z<0.7$.

This paper describes the photometric target selection, spectroscopic
observations and data analysis for the LRGs, and summarises the
properties of the sample.  Basic cosmological parameters are being
derived from the full sample, such as luminosity functions
\citep{dw06} and spatial correlation functions and clustering
\citep{nr06}.  \citet{ir06} investigate the incidence of star
formation as a function of redshift and have identified rare `k+a'
galaxies which had a substantial burst of star formation within the
last billion years, while \citet{ems06} identify and determine the
cosmic evolution of almost 400 2SLAQ LRGs which are catalogued radio
sources.  The 2SLAQ LRG redshifts have been used by \citet{pad05} and
by \citet{coll06} and \citet{bc06} to validate the determination of
SDSS photometric redshifts.

\section{Input catalogue and LRG sample definitions}

Distant LRGs were selected on the basis of SDSS photometric data (see
\citealt{fuk96} for the definition of the $ugriz$ $AB_{\nu}$ system),
using a two-colour plot of ($r-i$) against ($g-r$) and the $i$--band
magnitude.  Effectively, the selection uses a crude determination of
photometric redshift as the 4000\AA\ break moves through the $gri$
bands.  The colour criteria are similar to those used for the Cut II
sample of SDSS LRGs \citep{eis01}, although the selection here is
easier since beyond $z>0.4$ there is less confusion with
lower-redshift galaxies.  Targets were also required to have
non-stellar images.

The selection of targets into several priority classes was done
originally using the best SDSS photometry available in 2003.  The
later DR4 photometry \citep{ad06} has been used in the final redshift
lists and for the diagrams in this paper.  For most objects the
changes amount to at most a few hundredths of a magnitude, with
r.m.s. scatters of $\sim0.01$~mag and negligible mean shifts of only
$\sim0.001$~mag in each colour.  The photometry for some galaxies
changed by larger amounts, due mainly to changes in how composite
images were de-blended: as a result, a few tens of galaxies ($<~1$\%)
now appear to have colours inconsistent with their original selection
classification.

\subsection{Detailed selection criteria}

After some experimentation two main samples of LRGs were defined.  The
primary sample (Sample~8 in the data lists) has a surface density of
about 70~$\rm{deg}^{-2}$, chosen to maximise the completeness and
spatial uniformity of 2dF coverage for LRGs with $z>0.45$ (see Section
3.2).  The secondary Sample~9 consists of galaxies with $z\sim0.4$, to
utilise fibres which could not be assigned to primary targets.  A few
targets falling outside the selection boundaries were included as
`fillers' (Sample~0).

\begin{figure}
 \includegraphics[width=95mm]{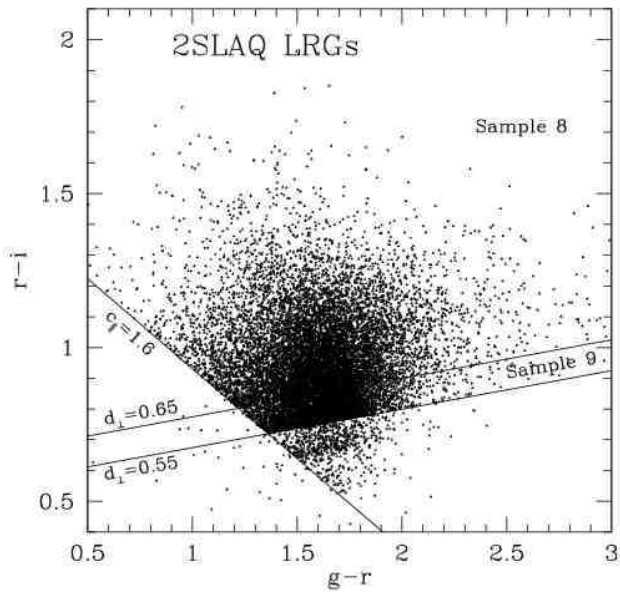}
 
 \caption{2SLAQ LRG selection boundaries in the $gri$ 2-colour
 diagram.  The primary (Sample~8) LRG targets lie in the top zone in
 the plot, above the diagonal lines $c_\parallel = 1.6$ and the upper
 $d_\perp$ cut at 0.65 (see text for details).  The secondary
 (Sample 9) objects lie between the two parallel $d_\perp$ cuts
 at 0.65 and 0.55.  Most of the points below the $d_\perp=0.55$
 cut (in the bottom right hand quadrilateral) represent lower
 redshift LRGs from the initial observing runs.
 Passively evolving LRGs with $z\sim0.55$ are expected to lie along an
 approximately vertical locus in this plot, with $g-r\sim1.7$ and with
 $r-i$ increasing from $\sim0.7$ to $\sim1.4$ as redshift increases
 from 0.4 to 0.7.  Nearly all lower redshift galaxies of all types lie
 below the $c_\parallel$ boundary in the bottom left hand corner.  The
 points with $(g-r)>2.3$ are mainly due to errors in the SDSS
 $g$-band photometry for such faint objects.}

 \label{2colsel}
\end{figure}

Fig.~\ref{2colsel} illustrates the colour selection boundaries
superimposed on the SDSS photometric data.  Most galaxies of all types
lie along a common locus in the lower left hand corner of this plot,
becoming redder in $g-r$ with increasing redshift until the 4000\AA\/
break moves into the $r$--band at $z=0.4$.  Thereafter, the $r-i$
colour becomes rapidly redder until the break moves into the $i$--band
at $z=0.7$.  Thus the most massive and luminous intermediate redshift
galaxies, i.e. LRGs with a dominant passively-evolving population, are
expected to lie along a vertical track with $g-r\sim1.7$ in
Fig.~\ref{2colsel}.

Cuts above lines of constant 
\begin{equation}
  d_\perp = (r-i) - (g-r)/8.0
\end{equation}
(cf. \citealt{eis01}) select early-type galaxies at increasingly high
redshift.  The top priority primary sample has $d_\perp \ge 0.65$
while the secondary sample has $0.65 > d_\perp \ge 0.55$.

A second cut with 
\begin{equation}
  c_\parallel = 0.7\times(g-r) + 1.2\times(r-i-0.18) \ge 1.6
\end{equation}
eliminates later-type galaxies.  The colours used here are the
extinction-corrected $modelMag$ colours: the definition of
these and other SDSS parameters can be found in \citet{sto02} or at
\verb"http://www.sdss.org/dr4/algorithms/photometry.html".  

The third main selection parameter is the de-reddened magnitude
in the $i$--band: the 2SLAQ LRGs have 
\begin{equation}
 17.5 \le i_{\rm deV} - A_i < 19.8
\end{equation}

where $i_{\rm deV}$ is the total magnitude based on a fit to a de
Vaucouleurs profile and $A_i$ is the extinction in the $i$-band.  This
choice of a fixed limiting magnitude enables the selection of bright
LRGs out to $z \sim0.8$, although it also means that the sample at $z
\sim0.5$ includes a substantial number of fainter early-type galaxies
with luminosity $\sim L^{\star}$.

Further cuts
\begin{equation}
 0.5 \le (g-r) < 3.0; \qquad (r-i)< 2.0
\end{equation} 
eliminate objects too far from the main LRG locus, probably mostly
composite objects or the result of photometric errors.

The photometric selection criteria are summarised in
Table~\ref{selcrit}.  The first two rows define the primary Sample~8
and secondary Sample~9, which comprise 67\% and 27\% respectively of
the final LRG survey.  4\% of the objects are in Samples~3--6, observed
in March and April 2003 using somewhat different criteria
(Section~2.2).  The final 2\% fell outside the selection
boundaries and have been assigned to Sample~0.

\begin{table}
 \caption{Photometric selection criteria.}
 \label{selcrit}
 \begin{tabular}{@{}ccccccl}
 \hline
Sample & $d_\perp$ & $c_\parallel$ & $i_{\rm deV}$ & $g-r$ & $r-i$ \\
  \hline
 S8 & $>0.65$ & $\ge1.6$ & $<19.8$ & $0.5-3.0$ & $<2.0$  \\
 S9 & $0.55-0.65$ & $\ge1.6$ & $<19.8$ & $0.5-3.0$ & $<2.0$ \\
\\
 S3 & $>0.55$ & $\ge1.6$ & $<19.5$ & $1.0-3.0$ & $<2.0$ \\
 S4 & $0.45-0.55$ & $\ge1.6$ & $<19.5$ & $1.0-3.0$ & $<2.0$ \\
 S5 & $0.25-0.45$ & $\ge1.6$ & $<19.5$ & $1.0-3.0$ & $<2.0$ \\
\\
 S6 & $>0.55$ & $\ge1.6$ & $<20.0$ & $1.0-3.0$ & $<2.0$ \\
  \hline
 \end{tabular}
\end{table}

Star/galaxy separation based on the SDSS images eliminates most
stellar contamination from the sample.  Two criteria were used,
\begin{equation}
 i_{\rm psf} - i_{\rm model} > 0.2 + 0.2\times(20.0-i_{\rm deV})
\end{equation} 
and
\begin{equation}
 radius_{\rm deV({\it i})} > 0.2
\end{equation}
but some cool M-dwarf stars remain and these comprise about 5\% of all
targets.  

Objects too diffuse to yield useful spectra using the 2~arcsec
diameter 2dF fibres were eliminated by requiring $i_{\rm fiber} <
21.2$ (such objects are liable to be spurious in any case).  The SDSS
3~arcsec fibre diameter is used for convenience, although the 2dF
fibres are only 2~arcsec in diameter.

The scatter of points on the red side of the main clump in
Fig.~\ref{2colsel}, beyond $(g-r)\sim2.3$, is mainly due to
photometric errors in the SDSS $g$-band.  Most points in this region
have $(g-i)\sim4.0$ and correspond to galaxies with $i\sim19.5$ and
hence $g>23$.  At this very faint level the mean error in $g$ is more
than 0.3~mag.

\subsection{Early observations}

Different cuts and priorities were used for the initial observing runs
in March and April 2003.  The original magnitude limit was $i_{\rm
deV} < 19.5$ but it became apparent that reliable redshifts could be
determined to somewhat fainter limits.  Tests showed that $i_{\rm deV}
< 19.8$ was a realistic limit in exposure times of 4~hours.  The
number counts are sufficiently steep that this relatively small change
had two very significant consequences: the proportion of high redshift
galaxies with $z>0.6$ doubled to $\sim15$\% and half of the primary
targets are in the faint extension.  Effectively the original limit
became the new median redshift, at $i_{\rm deV} = 19.48$.

Three colour-selected samples were defined originally: a top priority
class consisting of all objects with $d_\perp \ge 0.55$ and two
sparse-sampled classes with $0.45 \le d_\perp < 0.55$ and $0.25 \le
d_\perp < 0.45$ (Samples~3, 4 and 5 respectively).  These targets all
had $i_{\rm deV} < 19.5$ and $(g-r) \ge 1.0$.  A special faint
Sample~6 with $i_{\rm deV} < 20$ was observed in one test field, d10,
and was used to determine the optimum magnitude limit for the
remainder of the survey.  In addition to the different sample
specifications, there were subsequent small revisions to the SDSS DR1
photometry.  Most of the early observations could be reassigned to
Samples~8 or 9 and nearly all of the fields have been re-observed to
maintain the uniformity of the survey, but about 900 galaxies remain
outside the primary and secondary samples.

\section{Survey design}

The AAT 2dF fibre system \citep{lct02} is ideally suited to this
project.  The LRG surface density is $\sim$70 per sq deg down to a
magnitude limit of $i=19.8$, corresponding to an SDSS fibre magnitude
of $r_{\rm fiber}\sim22$.  2dF can provide reliable redshifts for over
90\% of such targets in four hours, in average conditions.
Atmospheric refraction limits the maximum time for which a given 2dF
configuration can be efficiently observed to about 3h for fields on
the equator, so most fields were observed on two consecutive nights.

The most practical survey design was two narrow strips along the
celestial equator, one in each Galactic hemisphere, given the early
SDSS photometric coverage and the requirement to work through complete
nights.  There was also a desire to overlap with previous surveys such
as 2QZ \citep{scr04} and the Millenium Galaxy Survey \citep{lis03,
spd05} and to enable follow-up observations from large telescopes in
both hemispheres.  The northern strip runs from 8.2h to 15.3h in RA,
broken into five sub-strips to utilise the best photometric data; the
southern strip runs from 20.6h to 4.0h.  Each strip can be covered by
a single row of overlapping 2dF fields.

\begin{figure}
 \includegraphics[width=100mm]{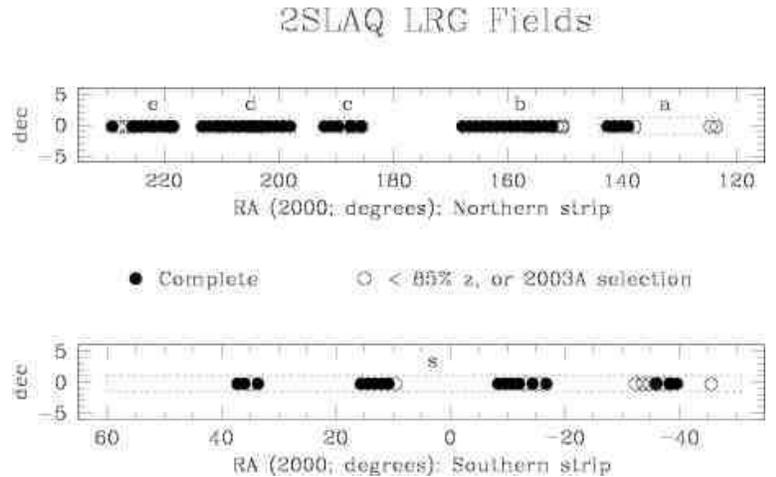}
 \caption{Layout of the 2dF 2SLAQ fields within the northern (upper)
 and southern (lower) Galactic strips.  Filled circles denote fully
 observed fields, i.e. with redshift completeness $\ge85$\%.  Open
 circles represent fields with completeness $<85$\% or with
 non-standard selection criteria.  The letters `a'-`e' and `s' signify
 the sub-strips and are used as the first character of the field
 names.}
 \label{fields}
\end{figure}

The ends of the strips come close to Galactic latitude $20\degr$ and
thus suffer from significant extinction of up to 0.4 mag in $g$ (based
on the maps of \citealt{sfd98}), plus substantial foreground star
contamination.  Priority was therefore given to fields nearer the
centre of each strip with $|b|>40\degr$ and extinction $A_g<0.2$ mag.
Fig.~\ref{fields} shows the layout of the target strips and the 2dF
fields actually observed.  There is considerable imbalance between the
two strips, with the northern strip containing more than two thirds of
the data and having much higher completeness.  This was an accidental
consequence of the nights scheduled and variable observing conditions.

\subsection{Tiling pattern}

The principal objective was to maximise the total number of objects in
a photometrically well-defined sample, with good coverage on different
spatial scales and minimal incompleteness.  For simplicity it was
decided to use a fixed field separation.  For the initial observations
in March and April 2003 the centres were set $1\degr$ apart but this
was increased to $1.2\degr$ for all subsequent observations, following
modelling tests of the yield.

Two strategies helped to maximise the fraction of targets which were
observed.  About 28\% of the targets lie in the overlap regions of
adjacent 2dF fields.  Targets which had been observed in the
configuration for one field were given lower priority in the adjacent
fields.  A sequence of alternate fields was observed first, followed
by the partially overlapping fields in a second pass along the strip.
This led to independent repeat observations for a few percent of the
total sample, giving a valuable check on the internal accuracy of the
data.  The second strategy arose because each field was observed on
two or more nights.  A significant fraction of the targets, mainly the
brighter galaxies and most of the M-stars, had good spectra and
unambiguous redshifts after only one night.  Typically some 10--20\%
of the fibres could then be re-allocated to new targets.  This helped
in crowded fields in particular, since due to physical constraints 2dF
fibres have a minimum separation of at least 30~arcsec.

\subsection{Target distribution across the field}

Tests of the 2dF {\sc configure} software showed that the algorithm
which assigns fibres to targets could introduce significant patterns
into the distribution of observed objects across the 2dF field.  The
main effect was a discontinuity in density at $0.25\degr$ radius, if
the mean number of targets per field in the input catalogue was much
higher than the number of fibres.  There could also be systematic
effects depending on the ordering of the input catalogue.  Both
effects were avoided in the 2SLAQ surveys by keeping the surface
density of the top priority targets below 70~$\rm{deg}^{-2}$ and by
randomising the order of the targets in the input catalogue.

\begin{figure}
 \includegraphics[width=95mm]{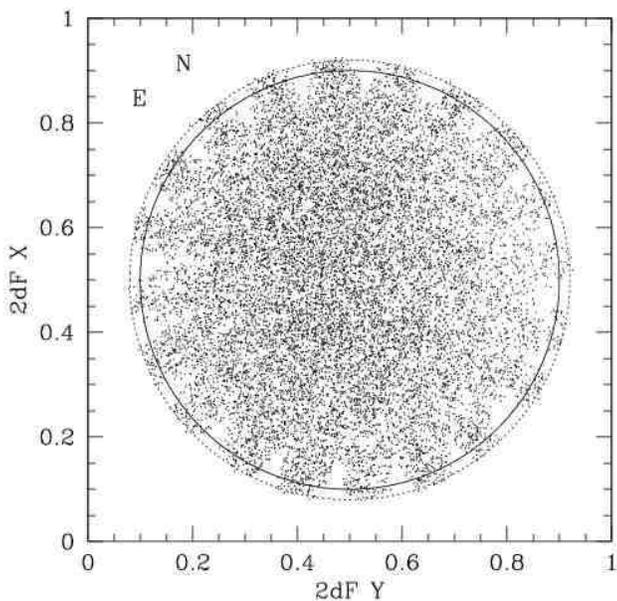} 

 \caption{The distribution of LRG targets across the 2dF field, with
 north to the top and east to the left.  The dotted circle represents
 the nominal $2.1\degr$ field diameter; the solid circle at $1\degr$
 radius indicates the effective area covered.  The empty triangles
 around the edge are inaccessible to the fibres feeding the single
 spectrograph used for the LRGs.  The lower density towards the
 right-hand side of the field is due to the allocation of 20--30 of
 the western fibres to sky: those fibres feed the central rows in the
 spectrograph camera.}

\label{xy}
\end{figure}

There are instrumental constraints on the targets accessible to 2dF.
The nominal $2\degr$ field can be up to $2.1\degr$ in diameter, but
this varied slightly during the survey due to both hardware and
software limits.  There are also 20 small triangular regions around
the edge of the field which are inaccessible to the fibres feeding one
spectrograph.  The net result is that the effective area of the survey
can be approximated by a row of overlapping $2\degr$-diameter circles.
Fig.~\ref{xy} shows the distribution of targets across the 2dF field,
for all objects in the final redshift catalogue.  There is a slightly
lower density of points towards the right hand edge of the field due
to the asymmetrical distribution of sky fibres (see Section~4.2).
Both this density gradient and the empty triangles on the eastern and
western edges are largely eliminated when the data from overlapping
fields are combined, but the northern and southern triangles remain.

\subsection{Biases in the targets observed}

In addition to the inaccessible regions shown in Fig.~\ref{xy}, each
fibre button has a large `footprint' with a long tail due to the
fibre, which runs across the front of the 2dF field plate.  It is
never possible to access targets within 30~arcsec of another which has
already had a fibre allocated, and in some directions the exclusion
zone is much larger.  There is therefore a strong intrinsic bias
against observing close pairs of galaxies, and very few fibres can be
allocated to members of one cluster.  This bias is partially overcome
within the overlap regions of adjacent 2dF fields, and a few fields
were independently observed in different runs.  The bias is also
alleviated by the policy of re-allocating some fibres after the first
night on a given field (Section~3.1).  However, there must remain a
substantial bias against close pairs of targets and the level of
incompleteness will have to be assessed by comparing the objects
actually observed with the full input catalogue.

There is a similar conflict between the LRGs and QSOs, since the two
classes of objects were observed simultaneously.  Thus, although one
of the strengths of 2SLAQ is having samples of LRGs and QSOs in the
same fields and within the same redshift range, there is a bias
against QSOs which lie near LRGs since the LRGs were given higher
priority than the QSOs in the allocation process.

\section{Observations}

\subsection{Spectrograph set-up}

The LRGs were observed with 2dF Spectrograph Number~2, using a 600
lines~mm$^{-1}$ $V$ grating.  The detector was a Tek1024 CCD with
$1024\times 1024$ pixels, giving a dispersion of 2.2\AA\ pixel$^{-1}$
and an effective resolution of about 5\AA\ or $R\sim1000$.  Almost all
the spectra were taken with the central wavelength set to 6150\AA,
covering the range 5050\AA\ to 7250\AA.  This enables secure
determination of redshifts for $z > 0.3$ using the Ca~II H\&K lines
and the detection of [O~II] 3727\AA\ down to $z = 0.35$.  For the
first observing run in March 2003 the central wavelength was set to
6350\AA\ but almost no useful data were obtained beyond 7250\AA, due
to telluric features which did not cancel well in the low dispersion
2dF spectra.  More details of the spectrographs and their performance
are given in \citet{lct02}.

\subsection{Sky fibres}

A subset of the 2dF fibres was assigned to clear sky positions, after
checking the SDSS images.  These are used to determine a median sky
spectrum which is subtracted from all target spectra.  For each frame
30 fibres were assigned to sky (later reduced to 20, after tests
showed that this resulted in negligible degradation).  These sky
fibres were selected to lie in the central half of the CCD
(i.e. between spectra 50 and 150) to keep the sky spectra uniform,
although this results in some mismatch to spectra at the edges of the
CCD due to optical distortions in the camera.  In principle it is
possible to correct for these but this has not yet been implemented in
the 2dF software.  The fibres which feed the centre of the CCD happen
to lie on the western side of the 2dF field, which causes the slight
density gradient seen in Fig.~\ref{xy}.

\subsection{Observing pattern}

The standard pattern was to take sets of seven exposures per field: a
quartz lamp flat-field, an arc lamp exposure using 4 Cu--Ar lamps and
2 He--Ar lamps, $4\times 1800$s exposures on the targets and a final
arc exposure.  Thus in good conditions four or occasionally five sets
of observations were obtained per night and four 2dF fields could be
fully observed in two nights.  In poor conditions the observations
were extended over extra nights, until the required mean
signal-to-noise per pixel (S/N) (and hence redshift completeness
fraction) was reached.

The 2SLAQ surveys were supported approximately equally by the
Australian and British time assignment panels, plus a few Director's
discretionary nights, to give a total allocation of 87 nights between
March 2003 and August 2005.  On average about 60\% of the time was
useable.  A total of 102 datasets was obtained for 80 2dF fields, 54
in the northern strip and 26 in the south.  Observations were deemed
to be complete (i.e. reliable redshifts for at least 85\% of the
primary targets) for 72 fields while 8 are either incomplete or used
non-standard selection criteria.  The final yield was very close to
one completed field per allocated night, about half of the best
possible rate.  This is lower than the overall fraction of clear time
because 2dF cannot be re-configured quickly.

The 80 distinct fields which were observed are illustrated in
Fig.~\ref{fields} and all the datasets are listed in Appendix~A, along
with various quality parameters which are discussed in Section~5.5.
Some fields have more than one dataset because data were obtained in
separate observing runs, sometimes with different parameters or target
selections, and were reduced independently.  There is some scope for
combining such datasets before analysis which would lead to better
quality spectra for repeated targets, but the gain in redshift
completeness will be very small.

Given that it was impossible to cover entire strips in the time
available, the aim was to observe several continuous sub-strips within
each Galactic hemisphere.  In the event, the time allocated and the
observing conditions led to more than twice as much data being
obtained in the northern strip as in the south.  Nine separate
sub-strips have been observed with varying levels of completeness.
The proportion of primary (Sample 8) input targets for which reliable
redshifts were obtained reaches 90\% in the best regions such as
northern sub-strips `b' and `d', each of which is $\sim17\degr$ long.
Each of these sub-strips covers a volume of space comparable to that
observed in the northern 2dFGRS strip, since the angular extent of the
latter was about five times larger but the mean redshift was only a
fifth as high.

\section{Data reduction and analysis}

\subsection{Extracting the spectra}

All of the raw data have been analysed using the AAO {\sc 2dfdr}
software \citep{bhc04}.  For each field, the location of the fibres on
the CCD was determined using a quartz lamp exposure which was also
used as a flat field to remove pixel-to-pixel sensitivity variations.
Two arc exposures provided wavelength calibration.  All spectra were
scaled according to the relative throughput of the fibres, as
determined from the strongest night sky lines, and a median sky
spectrum was subtracted from each object spectrum.  The different
frames for each field were combined using mean flux weighting, which
takes account of the variable signal levels arising from changes in
`seeing', transparency or exposure time; cosmic ray events were
removed during this final step.

The {\sc 2dfdr} software was developed for the analysis of the
2dFGRS/2QZ data.  For those surveys, the data for each field consisted
of several similar frames with precisely the same 200 targets, all
taken on the same night.  The {\sc 2dfdr} software was modified during
the course of the 2SLAQ surveys to cope with data taken on different
nights, sometimes with significant changes to the central wavelength
and often with altered allocations of fibres to targets.

\subsection{Redshift determination}

The LRG redshifts were derived using a modified version of the {\sc
Zcode} Fortran program, developed by \citet{wjs99} and others for the
2dFGRS spectra of low redshift galaxies \citep{cdm01}.  That program
determined two independent redshifts whenever possible, one based on
discrete emission lines and the other using cross-correlation with a
set of template spectra.  The higher resolution 2SLAQ LRG spectra
cover only half the wavelength range and the targets have been
photometrically selected to be predominantly passively evolving
early-type galaxies at much higher redshifts ($0.3 < z < 0.8$).  Thus
it is rarely possible to derive a secure redshift using emission lines
alone: the only line which is common is [O~II] at 3727\AA, detected by
the {\sc Zcode} in about 20\% of the spectra.  The 2SLAQ version of
the {\sc Zcode} derives redshifts by cross-correlating each spectrum
against a set of templates and uses any emission lines only as a check
on the cross-correlation results.

The cross-correlations are done using the method of \citet{td79},
which involves Fourier transformation of all the spectra and
templates.  The program selects the most plausible redshift for each
object and assigns a quality parameter `Q', which is a measure of the
consistency and reliability of the initial set of redshift estimates.
Another indicator of the reliability of the redshift is given by the
\citet{td79} parameter `$r$' (called $r_{\rm x}$ here to avoid
confusion with the photometric $r$), which measures the height of the
cross-correlation peak relative to the general noise level.  The
spectrum, cross-correlation function and redshift for each target are
shown on an interactive graphic display.  The operator can either
accept the automatic redshift or choose a different value (e.g. by
making alternative identifications of strong features or by selecting
a different template).  The operator assigns an independent quality
parameter to the reliability of each redshift.

The LRG version of the {\sc Zcode} can be run in several modes: the
basic fully interactive mode in which every spectrum is visually
checked; a quick mode in which only spectra with low quality redshifts
are inspected; and a fully automatic mode.  The advantage of
inspecting every spectrum is that unusual or extreme objects can be
identified, including composite spectra; the automatic mode has the
advantages of being more consistent and objective.  The semi-automatic
`quick' mode provides a good compromise between the two methods and is
normally used for initial data reductions at the telescope. 

Given the rather homogeneous 2SLAQ spectra, it has proved possible to
tune the automatic {\sc Zcode} to give very reliable results for most
objects.  Fewer than 1\% of the galaxies give significantly discrepant
redshifts from two independent analyses of the same dataset or from
repeat observations of the same targets (Section~6.1).  A majority
(72\%) of the redshifts for the initial data release have been derived
automatically, without visual inspection.  Some datasets from each
observing run have been re-analysed in fully interactive mode to
verify that the error rate remains low.

All redshifts have been corrected to be heliocentric.  Information on
the templates and other details specific to the redshifting of the
2SLAQ LRGs are given in Appendix~B.

\subsection{The redshift quality parameter $Q$}

The principal indicator of LRG redshift reliability is the parameter
$Q$.  Note that $Q$ does not necessarily indicate the quality of the
actual spectrum.  Although the two are strongly correlated in general,
there are cases where a very poor spectrum yields an unambiguous
redshift, and conversely.  Two quality values are output by the {\sc
Zcode}: the automatic value ($qa$) set by the code itself, and the
operator value ($qop$) set during the visual inspection of the data.
The preferred value is generally the operator value, where both exist.

The definitions of the operator values are as follows:

$Q=4$: A good spectrum with an unambiguous redshift (usually $r_{\rm x}
> 10$), based on several strong features.

$Q=3$: A single strong cross-correlation peak (with $r_{\rm x} > 5$) and
some obvious features (most often H\&K, G-band and/or Balmer lines).

$Q=2$: A plausible redshift but not completely convincing ($r_{\rm x} >
3.5$), or where alternative redshifts are possible.

$Q=1$: A spectrum where no believable redshift can be found.

In each case, the presence of [O~II] 3727\AA\/ emission (or, rarely,
[O~III] at 5007\AA) at the absorption redshift can be used to increase
the $Q$ value by 1 unit (but with $Q=4$ remaining the maximum for $qop$).
Checks on multiply-observed objects from overlapping fields, and from
repeat observations of the same sets of targets, show that the
reliability of the different classes is $>99$\% for $Q=4$,
$\approx95$\% for $Q=3$ and $\sim50$\% for $Q=2$ (see also
Section~6.1).

The automatic quality parameter $qa$ is derived somewhat differently
but produces very consistent results.  The evaluation is done in two
stages: first the cross-correlation results are compared for the full
set of templates to determine the parameter $qx$ for the most probable
redshift (i.e. the one which gives the highest $r_{\rm x}$ value), and
then the list of possible emission lines is checked for any
identifications at the same redshift.  The $qa$ value is also based on
the consistency of the redshift across the set of templates.  The
M-stars are dealt with slightly differently, since they only ever fit
one template (and that template, T8, never gives a correct redshift to
any galaxy).  $qa$ takes values in the range 0 to 5.

Comparisons between automatic and manual (i.e. fully interactive)
analyses of the same datasets show good agreement between the two sets
of $Q$ values.  In particular, it is very rare to get any discrepancies
between redshifts for which both methods give $Q \ge 3$, which
comprise 90\% of the targets for most datasets.  A few percent of
targets lie on the $Q=2-3$ borderline and these sometimes show
discrepant redshifts, as is to be expected.

\subsection{Quality of datasets}

The overall quality of each dataset (i.e. each set of combined frames
for a given 2dF field) can be evaluated in several ways, using four
global parameters listed in Table~\ref{fldlist} in Appendix~A.  These
parameters are the number of raw data frames, $N_\mathrm{exp}$; the
mean S/N per pixel; the mean difference between the logarithm of the
counts obtained and the expected magnitudes of the targets,
$D_\mathrm{mag}$, using here the SDSS $r_{fiber}$ parameter; and the
percentage of targets with reliable redshifts.  The redshift success
rate, defined as the fraction of targets which have redshift
reliability $Q$ of 3 or more, is the most important one for the LRG
survey, while the mean S/N should be the most reliable measure of
overall quality.  The observing strategy attempts to minimise the
variations in mean S/N by taking more frames in poor conditions, but
this is only partially successful since non-statistical errors begin
to become significant when many frames are combined.

\begin{figure}
 \includegraphics[width=84mm]{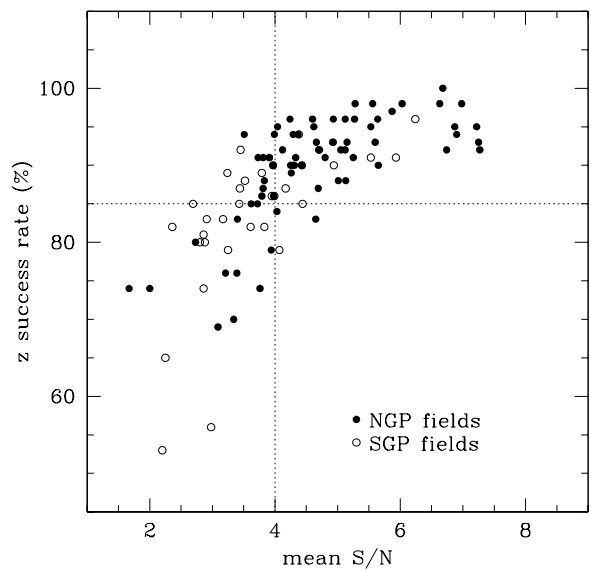}

 \caption{The redshift success rate for each dataset as a function of
 mean S/N per pixel per object.  The redshift completeness can fall
 catastrophically when the mean S/N is below 4.0 (vertical dotted
 line): the observations were repeated for most such datasets.  The
 horizontal dotted line marks the 85\% success rate used as the
 threshhold for deciding that the observations of a field were
 complete.}

 \label{yields}
\end{figure}

Fig.~\ref{yields} shows the correlation between $z$ success rate and
mean S/N (per pixel, per object) for all the datasets listed in
Table~\ref{fldlist}.  The northern Galactic strip has systematically
better data than the southern strip, mainly due to better weather
conditions.  Datasets for which $S/N > 5$ always have $z$ completeness
$>90$\% while $S/N > 4$ normally results in completeness of more than
85\%, the level chosen as the practical limit for deciding when to
terminate the observations of a field.  Most of the datasets with low
completeness were re-observed to give higher final redshift
completeness.  Data from fields with less than 85\% final completeness
are retained in the LRG catalogue but must be treated with caution in
any statistical studies.  Nearly all the datasets with S/N$>6$ are for
northern fields observed early in 2003, when a brighter magnitude
cut-off was used.

\subsection{Completeness and uniformity of the survey}

\begin{figure}
 \includegraphics[width=84mm]{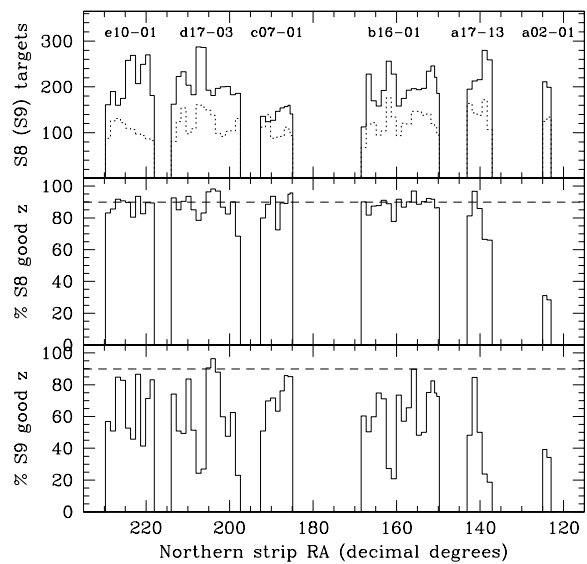}

 \caption{The completeness of data for targets in the northern strip.
 The top panel shows the number of potential targets in each 2dF
 field, for the primary Sample~8 targets (solid histogram) and the
 secondary Sample~9 targets (dotted histogram).  The middle panel
 shows the percentage of primary targets for which reliable redshifts
 were obtained (the horizontal dotted line marks 90\% overall
 completeness) while the bottom panel shows the same for the secondary
 sample.  The labels in the top panel give the names of the 2dF fields
 within each sub-strip.}  

 \label{completeness8}
\end{figure}

The 2SLAQ LRG survey is spatially incomplete (see Fig.~\ref{fields})
and has a variable redshift success rate (Table~\ref{fldlist}).
However, a particular effort has been made to maximise the
completeness for the primary targets within several continous
sub-strips.  The top panel of Fig.~\ref{completeness8} shows the
number of potential targets in each field of the northern strip for
Samples~8 and 9, with considerable cosmic variance between fields.
The middle panel shows the percentage of primary targets with reliable
redshifts, which is close to 90\% on average and reasonably uniform.
This is the overall completeness, i.e. the fraction of targets
observed multiplied by the redshift success rate.  The bottom panel of
Fig.~\ref{completeness8} gives the equivalent plot for the secondary
Sample~9 objects, for which the completeness is much lower and much
less uniform, ranging from 20\% up to 90\%.  Sample~9 utilised only
those 2dF fibres which could not be placed on primary targets: it
tends to be least complete in the highest density fields.

The completeness is particularly poor for the low Galactic latitude
fields a01 and a02, which were observed only with the original 2003
selection criteria.  Similarly, the southern field s01 was abandoned
after its high levels of stellar contamination and foreground
reddening were recognised.

It is difficult to define the precise boundaries of the 2SLAQ survey,
given the overlapping tiling pattern, the inaccessible triangles near
the edge of the field (Fig.~\ref{xy}) and the fibre re-allocation
observing strategy.  One simple way to derive a very complete sample
is to work in rectangles which lie completely within the patterns of
overlapping 2dF circles.  Two such inscribed rectangles can be drawn
in the well-observed northern sub-strips `b' and `d', containing over
30\% of the entire LRG survey.  The targetting completeness is 94.5\%
for the primary Sample~8 and the redshift completeness is 96.7\%,
giving an overall completeness of 91.4\%.

The best way to evaluate the completeness of the 2SLAQ survey, and to
calculate its effective total area, is to create a spatial mask
including all objects in the input lists which could in principle have
been accessed by 2dF from at least one of the observed field centres.
This has been done by \citet{dw06} and the resulting mask will be made
available as part of the survey data.

\subsection{Output lists}

Two \verb"ascii" listings are generated by the {\sc Zcode} for each
2dF field.  One, \verb"filenamez.rz", gives the target names,
positions, fibre numbers, redshift parameters and some photometric and
quality data, with one line for each spectrum analysed.  The second,
\verb"filenamez.zlog", gives much more data on the redshifting
process, including the results of cross-correlating every target
against each template and lists of strong emission lines found.  Some
overall statistics, such as the numbers of reliable redshifts and the
mean S/N of all the spectra, are also recorded.  The \verb"*z.rz"
files are concatenated and sorted to generate the final output
redshift lists for the LRG survey.

The target names follow the standard IAU convention, i.e. using the
truncated J2000 position in Right Ascension and declination.  These
are quoted to a precision of 0.1~arcsec, which ensures unambiguous
identification of the objects.  However, one drawback is that
sometimes the names can change slightly when the input SDSS
photometric data are revised, or when positions are converted between
arcsec (and seconds of time) and decimal degrees or radians.  The only
safe way to make cross-identifications with other catalogues is to
look for positional coincidences, rather than identical names.

Multiple observations were made for about 20\% of the targets, either
because they lay in the overlap regions of adjacent 2dF fields or
because entire fields were re-observed.  The {\sc 2dfdr} and {\sc
Zcode} software originally assumed that all observations for a given
dataset were taken using the same configuration of 200 fibres with the
same spectrograph set-up.  Thus two or more independent spectra were
obtained for many objects and two catalogues have been generated, one
listing every spectrum obtained and the other with a single entry for
each discrete object observed.  The latter, which is the basic
redshift catalogue, simply uses the best available redshift for each
object.  The selection is based on the redshift quality parameter $Q$,
or on mean S/N when two spectra give the same $Q$.

Better spectra could be obtained for some repeated objects by
combining the independent spectra from different observing runs.
Every spectrum should also be checked visually, to pick up any very
peculiar objects or ones where the automatic redshifting has missed
some obvious feature such as a composite spectrum.  However, this will
make very little difference to the actual redshifts: there should
simply be a small increase in the number which are reliable.  The
radio sources which comprise about 3.5\% of the 2SLAQ LRGs
\citep{ems06} have been used as a quality control sample and confirm
this expectation.

The redshift lists are held in the 2SLAQ team archive at the
University of Queensland, \verb"http://lrg.physics.uq.edu.au/" and can
be accessed via links at the University of Portsmouth
(\verb"http://www.2slaq.info"), at the AAO
(\verb"http://www.aao.gov.au") and elsewhere.  The archives also
contain the full sets of 2dF reduced data, the individual LRG spectra
and sets of SDSS postage stamp images.  A catalogue of the best
available redshift for every object observed is being released in
mid-2006, together with the basic SDSS photometric data and other
parameters.  The final full set of 2dF spectra and SDSS photometry
will be made publicly available once the data re-reductions are
complete.

\subsection{Examples of the spectra}

\begin{figure*}
\centering
\centerline{\psfig{file=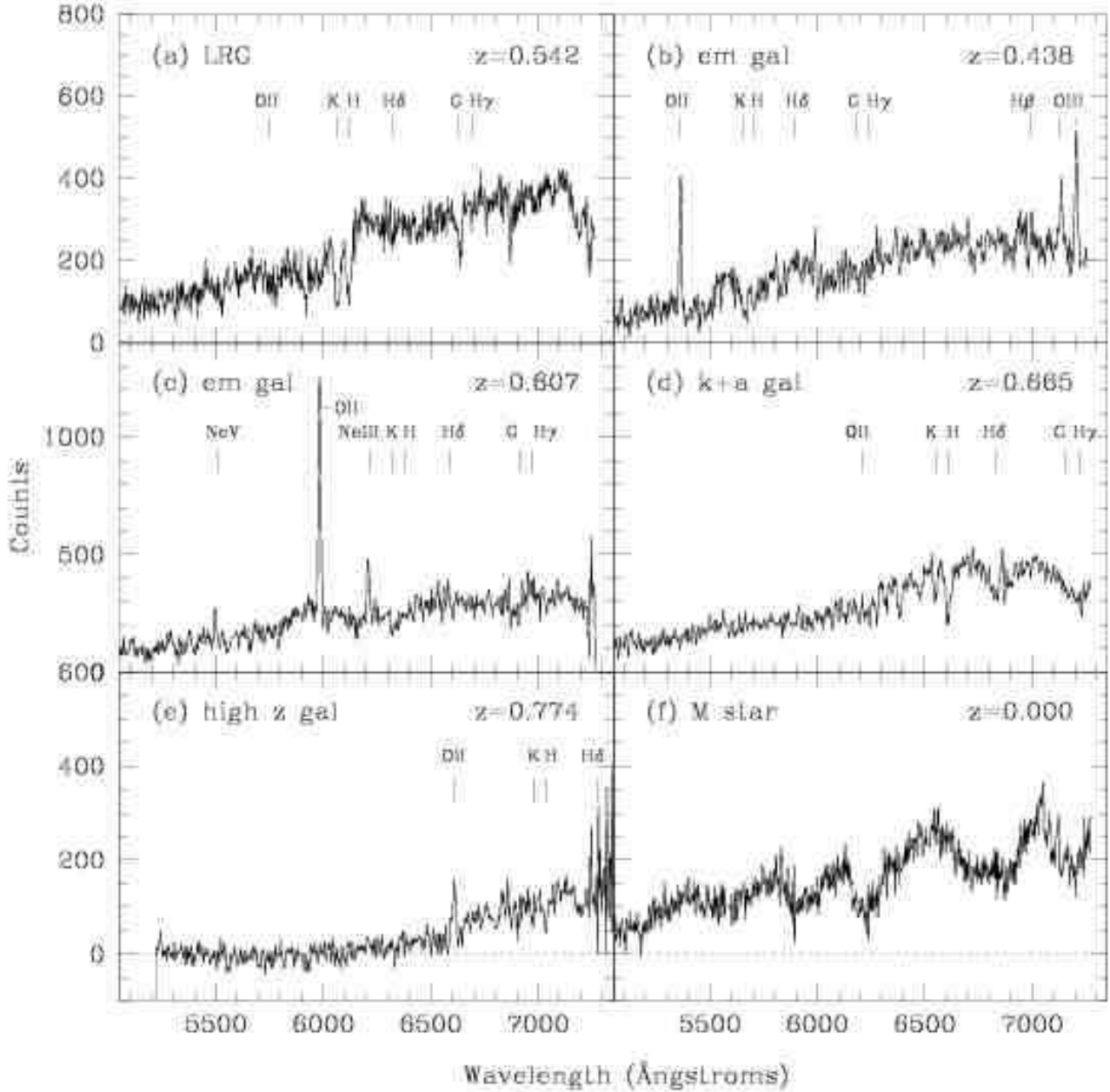,width=180mm}}

 \caption{Examples of 2SLAQ LRG spectra, with some major features
 identified.  No flux calibration has been applied.  The residuals
 from a few strong night sky lines have been excised, while the dips
 near 6850\AA\ and 7200\AA\ seen in several spectra are due to
 uncorrected atmospheric absorption bands.  Spectra (b) to (e) have
 been lightly smoothed.  The redshifts are indicated in the top right
 hand corner of each panel:
 (a) J110954.40+004843.1, a relatively high S/N spectrum of a typical
 LRG with redshift $\sim0.55$;
 (b) J022459.77$-$000156.4, a low redshift 2SLAQ galaxy with both
 [O~II] and [O~III] in emission;
 (c) J225439.77$-$001501.2, a rare emission line galaxy with higher
 excitation [Ne~III] and [Ne~V] as well as [O~II];
 (d) J212907.12+002405.4, one of the rare `k+a' LRGs with prominent
 Balmer lines in absorption and a weak 4000\AA\ break;
 (e) J122554.82$-$000752.4, one of the highest redshift 2SLAQ LRGs
 with $z = 0.774$;
 (f) J124853.73$-$010332.7, a typical foreground M-star.}

\label{spectra}
\end{figure*}

A number of representative galaxy spectra, plus one of the
contaminating M stars, are illustrated in Fig.~\ref{spectra}.

Fig.~\ref{spectra}(a) is a good spectrum of J110954.40+004843.1 at
$z=0.542$, a typical 2SLAQ LRG close to the median redshift.  The H\&K
lines of Ca~II and strong 4000\AA\ break are unambiguous determinants
of the redshift and several other stellar absorption features are very
obvious, but no emission lines are detectable.

(b) is the galaxy J022459.77-000156.4 at $z=0.438$, with both [O~II]
and the [O~III] doublet in emission.  This galaxy is close to the
lower redshift cut-off for 2SLAQ LRGs and shows why the spectra very
rarely yield reliable emission-line redshifts, since both [O~II] and
[O~III] fall within the 2SLAQ spectral window for only a very narrow
redshift range.

(c) is J225439.77-001501.2 at $z = 0.607$.  The best cross-correlation
used the K-type stellar template but the redshift reliability was low.
The redshift is secure only because high excitation emission lines of
[Ne~V] 3426\AA\ and [Ne~III] 3870\AA\ are clearly present.

(d) is J212907.12+002405.4 at $z = 0.665$, one of the relatively rare
`E+A' or `k+a' galaxies in the 2SLAQ sample (\citet{ir06}).
These galaxies show very strong Balmer absorption lines, with
H-$\epsilon$ making the Ca~II H~line apparently much stronger than K,
and a relatively weak G-band and 4000\AA\ break.  [O~II] 3727\AA\
emission is often present.  These galaxies are understood to have had
a substantial episode of star formation up to $5\times10^8$ years ago.

(e) is J122554.82-000752.4 at $z = 0.774$, one of the highest redshift
galaxies in the sample.  The continuum is close to zero over much of
the spectrum, but the strong H\&K lines plus [O~II] yield an
unambiguous redshift.  Like many high redshift LRGs, this galaxy shows
`k+a' features similar to Fig.~\ref{spectra}(d).  This spectrum was
obtained in March 2003 with the wavelength range shifted 200\AA\
redwards compared with all later spectra: it is apparent that the data
beyond 7250\AA\ were not useful.

(f) is J124853.73-010332.7 at $z=0.0$, a typical M-star with very
characteristic molecular absorption bands.  About 5\% of the targets
are foreground Galactic M-dwarfs; contamination by other stellar types
is negligible.

\section{Reliability and accuracy of the redshifts}

Considerable effort has gone into making the catalogue of 2SLAQ LRG
redshifts complete and reliable.  This involves not only determining
good redshifts for as many LRGs as possible but also identifying those
objects that are not LRGs within the redshift range of primary
interest.  Usually $\sim90$\% of the targets give unambiguous
redshifts in a first pass through the data.  By far the most common
cause of failure is low S/N, due to either target faintness or fibre
misplacement.  A few targets ($\sim0.1$\%) fail to give a reliable
redshift due to other causes such as instrumental artefacts or
proximity to bright stars, but these are not liable to introduce any
systematic bias into cosmological analyses of the LRGs.

About twenty composite spectra have been found and more would no doubt
be revealed by systematic searches: the {\sc Zcode} software finds
only the most probable redshift for each target.  Most of the
composites consist of an M-star plus a galaxy while a few show a
second emission line redshift on top of the primary spectrum.  The
photometric data are unlikely to be meaningful when a strong M-star
signature is present so the redshift is generally given as zero in
such cases.  Any secondary redshift is noted since it may still be
useful, e.g. if a target is identified as a radio source.

\subsection{Redshift reliability}

Several internal checks have been carried out by comparing different
analyses of the same data set, different sets for the same field, and
finally targets which have been observed independently in different
fibres from different configurations.

The best datasets yield very high redshift completeness, i.e. reliable
($Q \ge 3$) redshifts for $> 95$\% of the targets.  Comparisons
between analyses of such sets by different {\sc Zcode} operators show
excellent agreement in the preferred redshift, with only a handful of
objects differing, mostly among those with $Q=2$.  There can be more
spread in the $Q$ ($qop$) values since this is a somewhat subjective
measure, but nearly all agree within 1~unit.  

Several pairs of datasets for the same field have been compared.  The
number of significant redshift discrepancies, i.e. where two different
redshifts both with $Q>2$ are claimed, averages less than 1\%.
Most of these are cases where one or both cross-correlation redshifts
are on the borderline between quality 2 and 3, a few are due to the
automatic code missing an obvious emission line fit, a few are due to
manual operator errors, and a couple have been composite spectra with
two valid redshifts.

The largest set of independent spectra, mostly taken in different
configurations using different fibres, consists of repeat observations
in 2005 of 1253 objects from the first two years.  Just 9 pairs were
significantly discrepant, in that two apparently good redshifts with
$Q\ge3$ differed by more than 0.01.  Three of these were composite
star+galaxy spectra where both redshifts were valid and one was a very
unusual emission line galaxy.  The remaining five each had one
unambiguous redshift while the other was a marginal redshift close to
the boundary between $Q=3$ and $Q=2$.  Thus the error rate is only
about 0.5\%.

An external check of redshift reliability is provided by the SDSS LRG
Survey which has some overlap in the redshift range $0.3<z<0.5$ (see
Section~6.3).  The redshifts agree to within 0.002 for 143 of 145
galaxies in common, the only discrepancies being for 2 of the 3
galaxies for which the 2SLAQ redshift were unreliable, i.e. with $Q=2$.
The $Q\ge3$ redshifts are evidently 100\% reliable in this comparison
set.

\subsection{Significance of signal-to-noise ratio}

\begin{figure}
 \includegraphics[width=94mm]{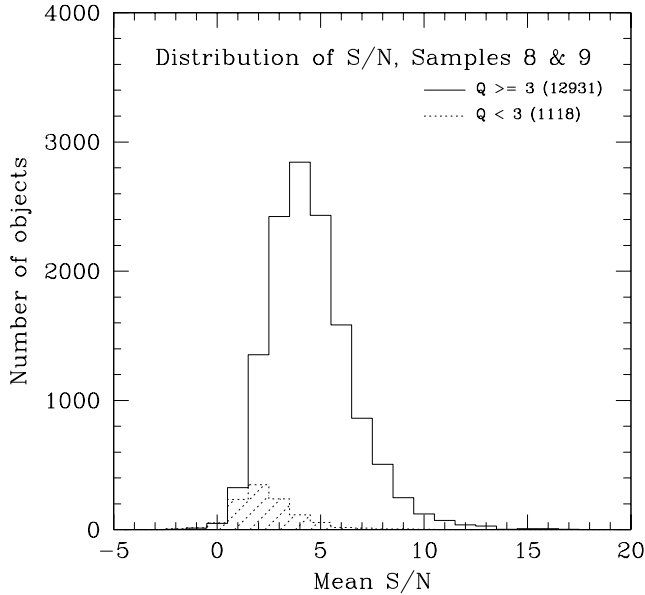}

 \caption{Histograms of mean S/N per pixel for all Sample~8 and 9
 spectra which gave reliable redshifts (solid line) and for those
 which failed (dotted line, shaded histogram).  95\% of spectra with
 S/N~$>2$ are adequate for determining redshifts.}

 \label{snhist}
\end{figure}

The redshift reliability depends strongly on S/N.  Histograms of S/N
for the objects with and without reliable redshifts are shown in
Fig.~\ref{snhist}.  The overall success rate (i.e. $Q\ge3$) for all
targets in Samples~8 and 9 was 92\%.  This rises to 99\% for spectra
with S/N~$>4$ but falls to just over 50\% for the $\sim5$\% of
spectra with S/N$<1.5$.

\begin{figure}
 \includegraphics[width=94mm]{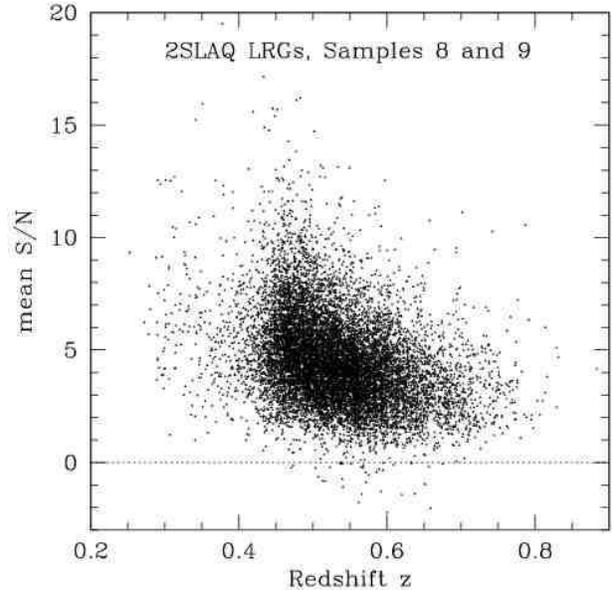}

 \caption{Mean S/N as a function of redshift $z$, for all primary
 (Sample~8) and secondary galaxies with reliable redshifts.  A few
 objects fall below the S/N~=~0 boundary (dotted line) due to errors
 in sky subtraction, but retain sufficient features for a reliable
 redshift determination.}

 \label{snz}
\end{figure}

Fig.~\ref{snz} shows the run of S/N per pixel with redshift, for
spectra which yielded reliable redshifts.  The average S/N decreases
as $z$ increases, from a median value of about 5.0 at $z=0.45$ to 3.0
at $z=0.65$.  This is partly because the galaxies become
systematically fainter with redshift and partly because only a small
section of spectrum redwards of the 4000\AA\ break remains within the
2SLAQ window.  Night sky emission lines and atmospheric absorption
features beyond 6800\AA\ exacerbate the latter effect.  The overall
shape of the distribution is a reflection of the plot of redshift
against magnitude (see Fig.~10, Section~8).  A few galaxies are listed
with negative S/N, which is clearly non-physical.  These are usually
spectra where the continuum drops below zero due to over-subtraction
of the sky or scattered light.  The redshift code can sometimes still
give a reliable $z$ value in such cases.

\subsection{Redshift accuracy}

The comparison with the SDSS LRGs provides a quantitative check on the
accuracy of both sets of redshifts.  There is a mean shift of 0.00031
between the two sets, arising from the use of air wavelengths for the
2SLAQ templates and vacuum wavelengths by the SDSS.  The rms
difference for the sample is 0.00040, after removing two discrepant
2SLAQ redshifts and a further three outliers with $dz\sim0.002$.  This
implies that the internal accuracy of the $Q\ge3$ 2SLAQ redshifts is
no worse than $\pm0.0003$.

A similar check can be derived from repeated 2SLAQ spectra.  The rms
scatter between sets of data taken in different years was typically
about 0.0006.  Some of this scatter arose from differences between the
templates, partly intrinsic and partly due to calibration errors.
When the velocity zero-points were corrected (Appendix~B1) the rms
scatter fell to 0.00045.  This again implies an internal error in a
single redshift of $\pm0.0003$.

\section{Survey statistics and coverage}

\subsection{Overall redshift statistics}

The overall statistics for the 2SLAQ LRG survey are summarised in
Table~\ref{zstats}.  Nearly 18500 spectra were obtained over three
years, for 14978 discrete objects.  13784 of these (92\%) have good
($Q>=3$) redshifts.  About 5\% of the targets turn out to be
foreground M~stars, leaving a total of 13121 galaxies of which 11451
have $z>0.45$.

These statistics are for all 80 discrete 2dF fields which were
observed.  A few fields have low quality data or were observed with
only the original photometric selection criteria.  In particular,
fields a01, a02, s01 and s12 have less than 50\% completeness for the
primary Sample~8 targets and should be ignored in statistical analyses
(see Table~\ref{fldlist}).

Three columns of figures are given in Table~\ref{zstats}.  Column~1 is
for the primary Sample~8 which is the most complete and homogeneous
sample, column~2 for the secondary Sample~9 and column~3 for all
objects observed.  Sample~9 is photometrically homogeneous with high
redshift completeness and only 1\% contamination by M-stars, but it
has very variable spatial completeness.

The first two rows of Table~\ref{zstats} give the number of discrete
objects observed and the number which have reliable redshifts.  Row~3
gives the number of contaminating stars (virtually all M-stars).
Row~4 is the number of {\it bona fide} LRGs and row~5 gives their
median redshift.  The final three rows give the numbers of galaxies in
the primary redshift target range of $0.45<z<0.7$ and in the low and
high redshift tails.  Most of the $z<0.45$ galaxies are either in
Sample~9 or were observed in the first semester, before the selection
criteria were refined.  Virtually all of the highest redshift
galaxies, with $z>0.7$, are from Sample~8.

\begin{table}
 \caption{Target statistics.}
 \label{zstats}
 \begin{tabular}{@{}lrrr}
  \hline
  Selection & Sample 8 & Sample 9 & All \\
  \hline		 
  observed     	&10072 & 3977 & 14978 \\
  $z$ ($Q\ge3$) & 9307 & 3624 & 13784 \\
  \\		 	              
  stars       	&  551 &   37 &   663 \\
  LRGs        	& 8756 & 3587 & 13121 \\
  median $z$    & 0.55 & 0.47 & 0.522 \\
  \\		 	              
  $0.45<z<0.7$  & 8289 & 2647 & 11196 \\
  $z < 0.45$    &  214 &  935 &  1664 \\
  $z > 0.7$     &  253 &    5 &   261 \\
  \hline
 \end{tabular}

\medskip
\end{table}

\subsection{Spatial coverage of the survey}

The total area covered by the 2SLAQ survey is approximately
180~$\rm{deg}^2$, calculated as the number of fields observed
multiplied by the effective area of each field, corrected for edge
effects and the overlap between adjacent fields.  However, this is not
a particularly useful number since the completeness of the survey
varies significantly from field to field and between the different
samples in each hemisphere.  Further complications arise from the
constraints on placing 2dF fibres in close proximity, mentioned in
Section~3.3.

A more useful statistic for comparison with other surveys is the
effective area of the survey, defined as the total number of targets
with reliable redshifts divided by their mean density in the input
catalogues.  This gives an effective area of approximately
135~$\rm{deg}^2$ for the Sample~8 targets and 90~$\rm{deg}^2$ for
Sample~9.  A more careful calculation by \citet{ems06}, using the
detailed survey mask of \citet{dw06}, revises these areas to
141.7~$\rm{deg}^2$ and 93.5~$\rm{deg}^2$ for Samples~8 and 9
respectively.

These figures imply that the overall completeness of the primary
sample is only about 75\%, falling to 50\% for the secondary sample.
However, as noted in section~5.6, the completeness of the primary
sample rises to $\sim90$\% in the best-observed regions.

\section{Properties of the LRG sample}

\subsection{Redshift distributions}

\begin{figure}
 \includegraphics[width=84mm]{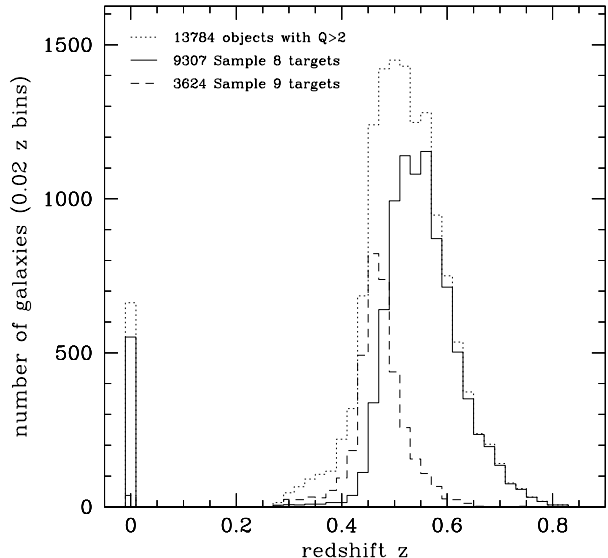}

 \caption{Redshift histograms for the 2SLAQ LRGs.  The solid line is
 for those objects in the primary sample (8) and the dashed line is
 for the secondary sample (9).  The histogram for all objects with
 reliable redshifts is shown as a dotted line.  The spike at $z=0$
 represents the 5\% contamination by foreground M-stars.  The
 effectiveness of the 2SLAQ photometric selection for $z\sim0.5$
 galaxies is evident.}  
 \label{zhist}
\end{figure}

A histogram of the redshifts of all the 2SLAQ galaxies is shown in
Fig.~\ref{zhist}.  It is obvious that the SDSS colour and magnitude
selection criteria produce a very clean sample of high redshift
galaxies.  Almost all the targets have $0.3 < z < 0.8$, apart from
$\sim5$\% contamination by foreground M-stars.  The dotted line is for
the entire set of 2SLAQ LRGs, including the early samples with
somewhat different selection criteria which contribute most of the
low-redshift tail.  The solid line is for the primary sample which has
median redshift 0.55 and 80\% of the redshifts within $0.45<z<0.7$, as
desired.  The sharp lower cut-off is set by the colour selection while
the declining tail for $z > 0.6$ is due to the $i < 19.8$ magnitude
limit.  The secondary (Sample~9) galaxies, shown by a dashed line,
have median redshift 0.47.

\begin{figure}
 \includegraphics[width=84mm]{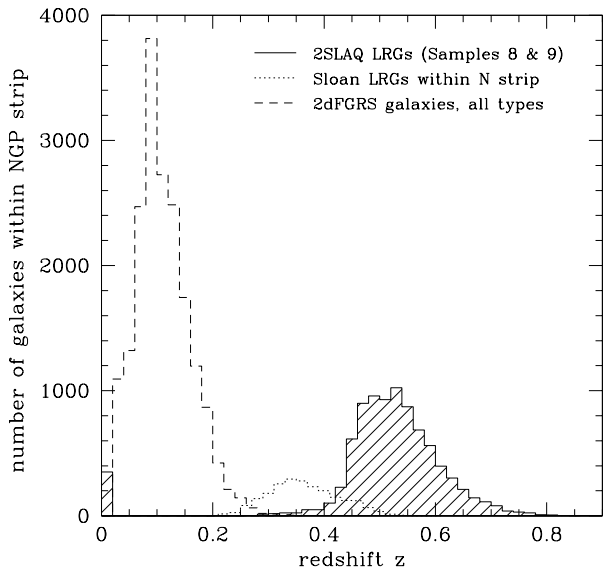}
 \caption{The redshift distribution of three surveys in the same
 $2\degr$-wide Northern equatorial strip.  The 2SLAQ LRG redshift
 distribution (shaded histogram) is compared with that of the SDSS LRG
 survey (dotted line) and galaxies of all types in the 2dFGRS (dashed
 line).}
 \label{zhistcomp}
\end{figure}

Fig.~\ref{zhistcomp} compares the redshift distribution of the 2SLAQ
LRGs with that of the SDSS LRGs \citep{eis01} and the 2dFGRS galaxies
of all types \citep{cdm01}, in each case counting only the galaxies
lying within the same $2\degr$ wide northern equatorial strip.  The
three surveys are complementary and together give good redshift
coverage from $z=0.02$ to $z=0.7$.  There is significant overlap
between the 2SLAQ and SDSS LRG samples at $z = 0.4$ but 2SLAQ provides
a much higher space density of galaxies with $z > 0.4$ than any other
current survey, as is needed for mapping the 3--D structure.

\begin{figure}
 \includegraphics[width=95mm]{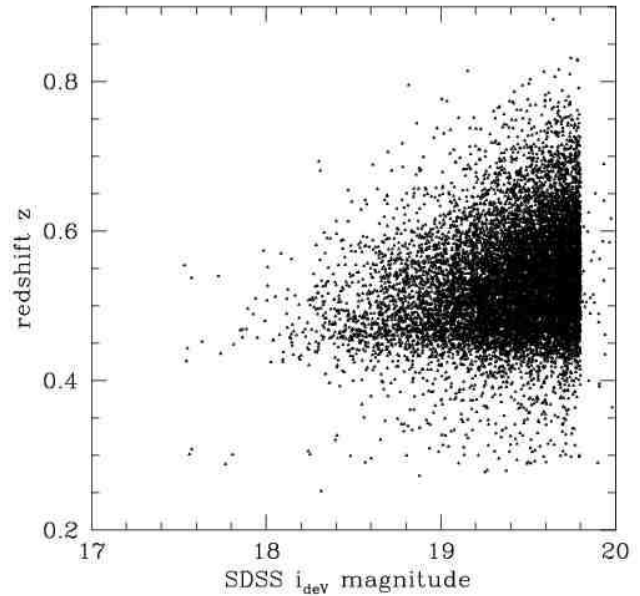}

 \caption{A plot of redshift against $i$-band magnitude for 11,000
 2SLAQ galaxies (Samples~8 and 9).  The upper diagonal boundary
 represents the most luminous galaxies at each redshift; the lower
 boundary at $z\sim0.45$ is set by the colour selection criteria.}

 \label{zi89}
\end{figure}

A plot of redshift against SDSS $i_{\rm deV}$ magnitude
(Fig.~\ref{zi89}) illustrates clearly the importance of a faint
$i$--band limiting magnitude in selecting high $z$ LRGs.  The number
of galaxies increases with increasing $i$ magnitude at all redshifts.
At the highest redshifts only a small number of intrinsically very
luminous galaxies are observed, while at lower redshifts the sample
includes early-type galaxies with luminosity $\sim L^{\star}$,
somewhat fainter than the `classical' definition of an LRG as having
$L \ge 3L^{\star}$ \citep{eis01}.  The sharp lower boundary near
$z=0.45$ demonstrates again the high efficiency of the SDSS
photometric selection criteria.

\subsection{2-colour distributions for different redshift ranges}

\begin{figure}
 \includegraphics[width=84mm]{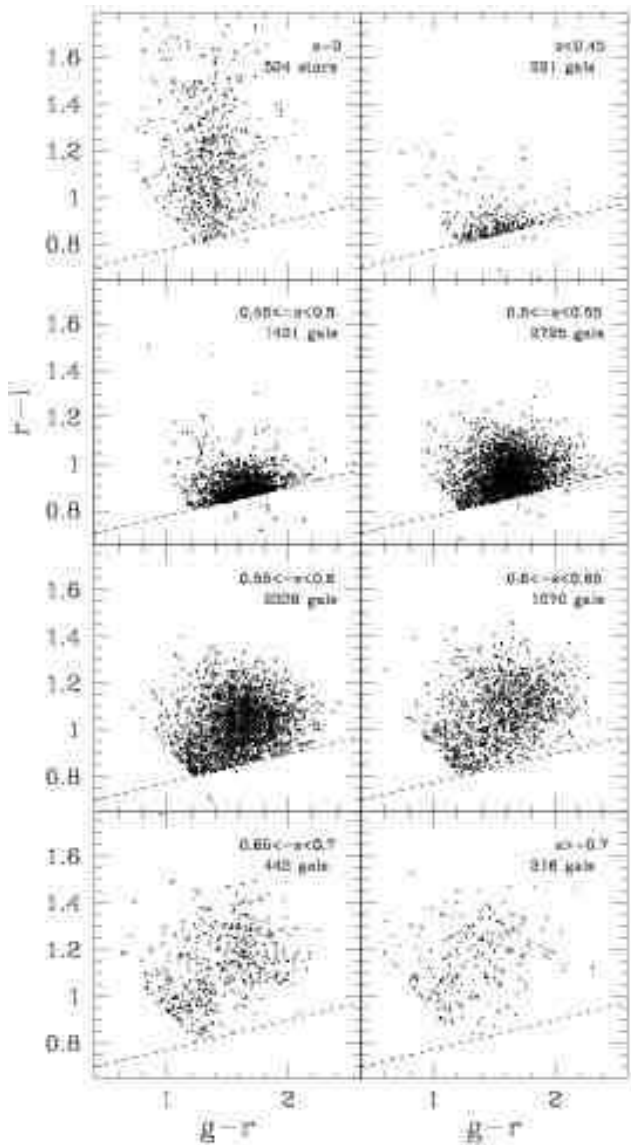}

 \caption{The distribution of LRGs in the 2-colour $(r-i)$ {\it
 versus} $(g-r)$ diagram as a function of $z$, for the primary
 (Sample~8) objects.  Each panel is for a successively larger redshift
 range, starting with the M-type stars at top left and ending with the
 highest redshift objects at bottom right.  The dotted line in each
 panel is the $d_\perp=0.65$ lower boundary to Sample~8, for
 reference (a few points appear below this line due to revisions of
 the photometry).  The main clump of LRGs moves to redder $r-i$ colours at
 approximately constant $g-r$ as $z$ increases, as the 4000\AA\ break
 moves through the $r$--band.}

 \label{2col8multi}
\end{figure}

The colour distribution of the targets varies strongly as a function
of redshift.  This is illustrated in Fig.~\ref{2col8multi} for the
primary (Sample~8) data.  Each panel is for a different redshift bin
in steps of 0.05 in $z$, apart from the low and high-$z$ tails.  

The top left panel shows the M-stars, which mostly occupy a vertical
strip with $g-r\sim1.4$ and extending to very red $r-i$ colours.  For
the galaxies in the remaining seven panels, the centroid of the main
concentration of points moves to higher values of $(r-i)$ at
approximately constant $(g-r)$ as redshift increases, due to the
4000\AA\ break moving through the $r$--band.  For $z>0.55$ the main
concentration of LRGs moves clear of the colour selection boundaries,
indicating that the primary LRG sample has high completeness above the
$i$--band apparent magnitude limit.  Careful analysis of the numbers
of SDSS and 2SLAQ LRGs \citep{dw06} shows they have almost identical
luminosity functions at $z=0.24$ and $z=0.6$, indicating that most
LRGs were formed at higher redshift and are evolving passively.  This
conclusion is confirmed by the composite LRG spectra for different
redshift ranges presented by \citet{ir06}: these are all very similar
to the spectrum of a nearby giant elliptical galaxy such as NGC~3379
(Template~2, Fig.~\ref{tempspec}) and show only slight evolution for
$0.2<z<0.7$.

The 2-colour distribution in Fig.~\ref{2col8multi} becomes more
diffuse as $z$ increases, which may be an indication of increasing
levels of current or recent star formation affecting the colours at
higher redshifts.  However, the errors in the colours also become
significant for the faintest and reddest objects, especially in the
$g$-band.  Objects with $\sigma_g>0.35$ have not been plotted,
following the discussion of photometric errors in Section~2.1.  Such a
cut eliminates 3\% of the primary sample, including virtually all
objects with $(g-r)>2.3$.

\subsection{The highest redshift galaxies}

\begin{figure}
 \includegraphics[width=84mm]{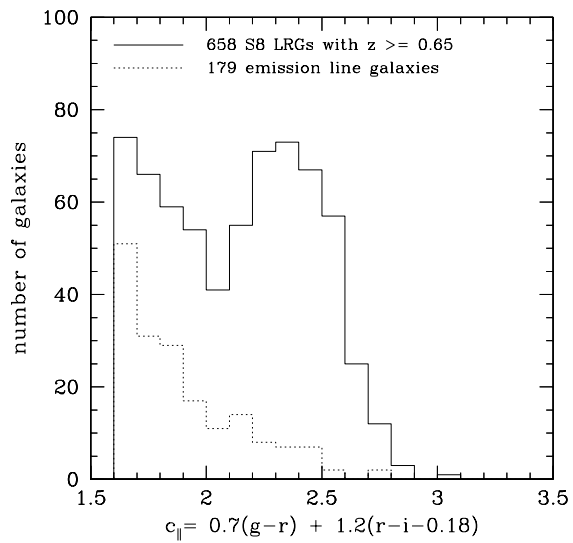}

 \caption{A histogram of the distribution of the $c_\parallel$
 photometric parameter for galaxies in the primary LRG sample with
 $z\ge0.65$ and $\sigma_g<0.35$.  The full line is for all LRGs, the
 dotted line is for those with [O~II] emission.  The sharp boundary at
 $c_\parallel=1.6$ marks the LRG colour selection cut-off.  The bins
 are independent, supporting the reality of the dip in the full
 distribution at $c_\parallel=2$.  The highly skewed distribution of
 emission line galaxies indicates the existence of two separate
 populations of high redshift galaxies, with only the redder peak
 corresponding to true LRGs.}

 \label{cparhist}
\end{figure}

For $z>0.55$ the colour distribution appears to be bi-modal, with a
clump centred near $(g-r) = 1.7$ and a second population towards the
lower left corner of each plot, truncated on the blue side by the
$c_\parallel > 1.6$ diagonal limit (cf. Fig.~\ref{2colsel}).  This
bimodality is shown more clearly in Fig.~\ref{cparhist}, a histogram
of the distribution of $c_\parallel$ for galaxies in the two highest
redshift bins from Fig.~\ref{2col8multi}.  Further evidence for the
reality of this division comes from the distribution of galaxies with
strong [O~II] emission, as detected during the redshifting process and
indicated by the dotted histogram in Fig.~\ref{cparhist}.  The em-line
galaxies are strongly concentrated towards the blue boundary.  The
equivalent widths of the emission lines are not large enough to cause
the colour shift directly in true LRGs: it appears most likely that
the `bluer' objects represent later-type galaxies with recent star
formation, spilling into the 2SLAQ sample from below the colour
selection boundaries.  \citet{ir06} and \citet{dw06} discuss this
effect in some detail, since it has to be corrected for in determining
the luminosity functions of the LRGs and in measuring the evolution in
the rate of star formation with redshift.

The fraction of emission line galaxies becomes particularly high for
the highest redshift group, with $z > 0.7$.  This is evidence for
higher rates of activity (either star-forming or AGN) at earlier
epochs, although the statistics have to be corrected for observational
selection since the useful spectral range is short and affected by
telluric features (see Fig.~\ref{spectra}(e)).  Sometimes such
galaxies can only be assigned reliable redshifts when [O~II] 3727\AA\
is present, so there is a bias towards recognising the more active
galaxies at high redshift.

The template statistics given in Appendix~B support the identification
of two types of galaxy among the highest redshift 2SLAQ LRGs.  The
main (redder) concentration of galaxies nearly all fit the composite
LRG spectrum (T1) best and must be true LRGs, while many of the bluer
galaxies lying close to the $c_\parallel$ boundary are best fit by the
later-type emission-line galaxy NGC~5248 (T3) (see Fig.~\ref{tzhist}).
The appearance of emission-line or star forming galaxies along the
$c_\parallel$ boundary is evident in the lower redshift bins as well,
at least down to $z > 0.4$, which probably indicates that the two
basic populations occur throughout the present sample as in other
galaxy redshift surveys.  This population is a contaminant in the LRG
sample, and has to be distinguished from the subset of true LRGs which
have emission lines due to either transient star formation or AGN
activity, as discussed by \citet{ir06}.

\subsection{The `failed redshift' objects}

\begin{figure}
 \includegraphics[width=84mm]{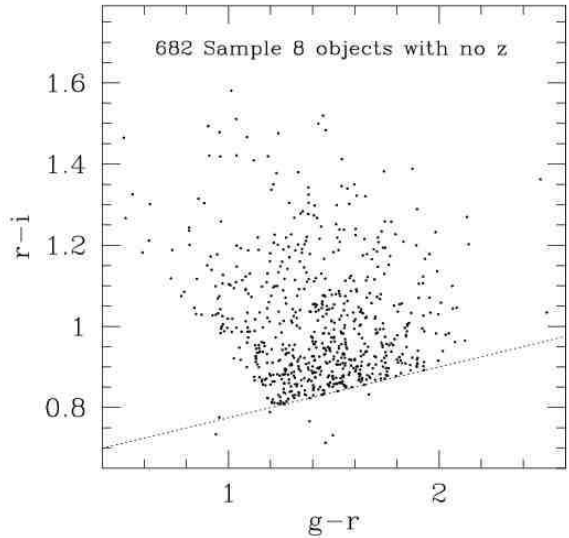}

 \caption{The 2-colour distribution of the 8\% of primary (Sample~8)
 LRGs with unreliable redshifts, i.e. with redshift quality $Q<3$
 (again omitting data with $\sigma_g>0.35$).  This distribution is
 significantly different from that of the full sample or any of the
 redshift sub-samples plotted in Fig.~\ref{2col8multi}.  A relatively
 high proportion of objects with extreme colours fail to give good
 redshifts, while {\it bona fide} LRGs in the clump with $z\sim0.55$
 are under-represented.  The `failed redshift' objects include a
 mixture of high redshift galaxies, galaxies with peculiar spectra,
 composite objects, lower redshift galaxies which are not LRGs, and
 LRGs which simply had weak spectra because of bad fibre placement or
 low throughput.}

 \label{noz}
\end{figure}

The two-colour distribution of the primary targets which were observed
but failed to yield a reliable redshift is shown in Fig.~\ref{noz}.
This distribution is not the same as for the full input sample
(Fig.~\ref{2colsel}) or any of the redshift subsamples in
Fig.~\ref{2col8multi}.  Evidently the redshift failure rate is a
complex function of redshift itself, and of other parameters such as
magnitude and colour.  Fig.~\ref{noz} suggests that a considerable
fraction of the `failed' objects are high redshift galaxies with $z >
0.7$, while there is also a significant population near the
$d_\perp=0.6$ cut-off with colours similar to $z < 0.4$ galaxies.
There are few failed redshift targets within the main clump of
$z\sim0.55$ LRGs seen in the central panels of Fig.~\ref{2col8multi},
despite these being the dominant population in the sample, indicating
that the identification of the true passive LRGs is very nearly
complete.

The overall distribution of the failed redshifts shows no significant
dependence on apparent $i$-band magnitude or on the general quality of
the datasets, as measured by redshift completeness or mean S/N
(Section~5.4).  Quantitatively, the fraction of failed redshift
objects with colours within the main concentration of LRGs (taken here
as the objects with $1.4<g-r<1.8, 0.9<r-i<1.2$) is only 5\%
compared with 8\% for all objects in Sample~8.  By contrast, 12\%
of the reddest objects with $r-i>1.2$ fail to yield reliable
redshifts and this fraction rises to 17\% for the bluest objects
with $g-r<1.2$.  

These variations demonstrate that there are real correlated
differences in colour and spectral type within the primary sample:
they cannot simply be due to either poor spectra or photometric
errors.  Many of the reddest and bluest objects must have spectra
which are not so well matched to any of the template spectra as are
the bulk of the LRGs.  Some may be active galaxies of various types,
some may be composite spectra due to chance alignments.  The `failed
redshift' list may also include some very high redshift objects with
$z\ge0.8$, given that such objects are difficult to recognise in the
2SLAQ data.  A few spectra with high S/N also fail to yield redshifts;
in at least one case this was due to contamination of the spectrum by
light from a nearby bright star with an almost featureless spectrum in
the 2SLAQ wavelength range.

\subsection{Completeness of the 2SLAQ samples}

The discussions of the colour distributions above, of sample
completeness in Section~5.5 and of redshift reliability in Section~6,
together indicate that the 2SLAQ sample of true passively evolving
LRGs is over 95\% spectroscopically complete and reliable for
redshifts between 0.5 and 0.65.  Few LRGs can have been missed, and
few non-LRGs included, down to the $i_{\rm deV} = 19.8$ magnitude
limit.  The primary Sample~8 is also almost spatially complete in the
most fully observed regions, assuming that the SDSS input catalogue is
itself almost complete, so that it should be possible to define a
sample of LRGs with an absolute completeness of better than 90\%.  

The primary Sample~8 2SLAQ LRGs therefore provide a good sample of
co-moving test particles for cosmological purposes and for
constraining LRG evolution.  In particular, most of the LRGs seen at
redshift 0.65 must be the progenitors of LRGs at $z=0.5$, and of the
passively evolving LRGs seen at lower redshift in other surveys.

The secondary Sample~9 2SLAQ LRGs are much less useful for studying
large scale structure, since their spatial completeness is very
variable from field to field.  However, this sample has a much tighter
colour distribution than Sample~8 (Fig.~\ref{2colsel}) and so
comprises a more homogeneous set of galaxies, with high redshift
completeness and little contamination, which will be very useful for
studying the evolution of LRGs.

\subsection{The spatial distribution of 2SLAQ LRGs}

\begin{figure*}
\centering
\centerline{\psfig{file=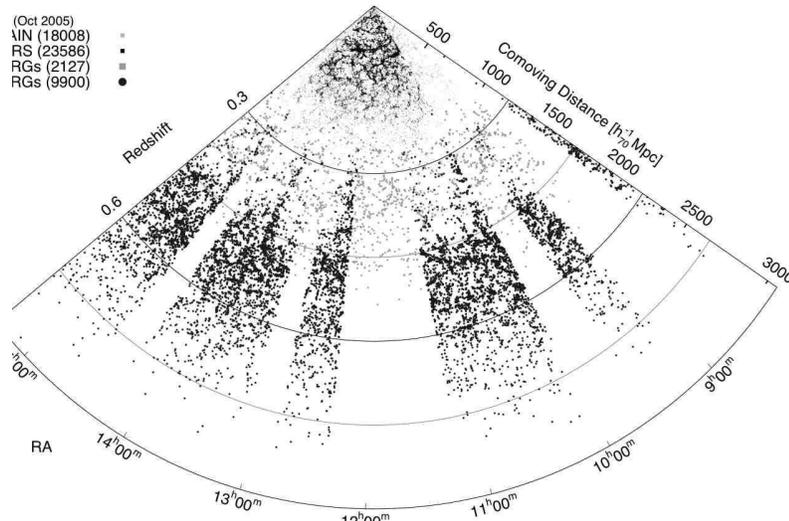,width=105mm}}

 \vspace{-6cm}

 \caption{A wedge diagram plotting redshift or distance as a function
 of RA for all the 2SLAQ LRGs in the Northern Galactic strip (large
 black points in discrete sectors).  Very small points with $z<0.3$
 show the 2dFGRS and SDSS MAIN galaxies of all types and the grey
 points at intermediate redshifts are from the SDSS LRG survey, all
 lying within the same $2\degr$-wide strip.  Considerable structure is
 evident within the LRG samples, similar to what is seen at much lower
 redshifts.}

 \label{PW_wedge}
\end{figure*}

A map of the spatial distribution of the LRGs within the more complete
northern slice is shown in Fig.~\ref{PW_wedge}, a plot of redshift (or
co-moving radial distance, assuming $\Omega_m=0.3$,
$\Omega_\Lambda=0.7$ and $H_0 = 70$\,km\,s$^{-1}$\,Mpc$^{-1}$) against
RA along the equatorial strip.  The 2SLAQ LRGs are plotted as large
black dots and lie in discrete sectors corresponding to the sub-strips
with the best SDSS photometry.  The distribution appears non-random
with higher density clumps and filaments and lower density voids.  The
thin pencil-beam extensions down to $z=0.3$ come from the 2dF fields
observed early in 2003, using lower-redshift selection criteria.  The
grey points with $0.25<z<0.5$ are the SDSS LRGs which lie within the
same $2\degr$-wide equatorial strip \citep{eis01} and the fine
structure with $z<0.2$ comes from a combination of the SDSS MAIN
\citep{str02} and 2dFGRS \citep{cdm01} galaxies of all types, also
within the same narrow strip.  The apparently finer structure seen at
low redshifts is mainly an artefact of the much higher surface density
of points available.  This wedge plot gives an excellent impression of
how the four surveys complement each other in terms of redshift and
spatial coverage.  The redshift histogram shown previously as
Fig.~\ref{zhistcomp} is effectively a projection of
Fig.~\ref{PW_wedge} along the redshift axis.
 
\begin{figure}
 \includegraphics[width=84mm]{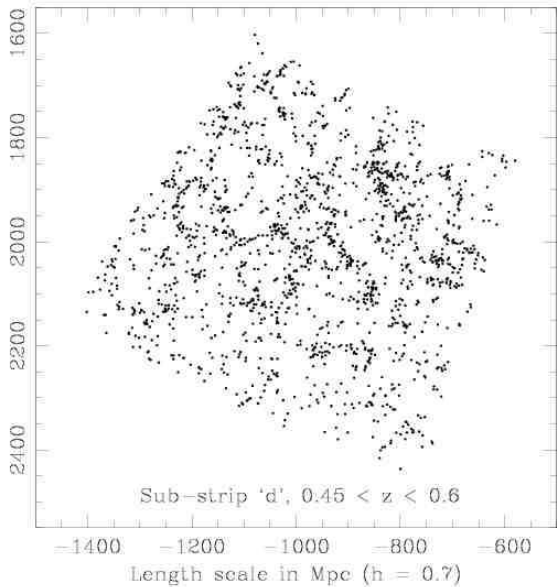}
 \caption{An enlarged portion of Fig.~\ref{PW_wedge} for sub-strip `d'
 at RA$\sim14$h.  This shows the spatial distribution of 1737 LRGs in
 the primary and secondary 2SLAQ samples (Samples~8 and 9) with
 $0.45<z<0.6$.  A pattern of voids and super-clusters linked by
 filaments or walls is apparent, on scales of hundreds of Mpc.}

 \label{wedge_d89}
\end{figure}

The large-scale structure in the 2SLAQ LRGs can be seen more clearly
at higher magnification.  Fig.~\ref{wedge_d89} is an enlargement of
part of Fig.~\ref{PW_wedge} corresponding to sub-strip `d', one of the
largest areas with contiguous coverage.  A pattern of voids and
super-clusters linked by filaments or walls becomes apparent, on the
same $\sim100$ Mpc scales as seen at $z\sim0.1$ in surveys such as the
2dFGRS \citep{pim04} and SDSS MAIN survey \citep{pb05}.

\section{Summary}

The primary objective of the 2SLAQ LRG survey, to obtain redshifts for
more than 10,000 galaxies with $0.45<z<0.7$, has been achieved.  The
galaxies lie in two thin sectors or wedges, one in each Galactic
hemisphere, corresponding to narrow strips running along the celestial
equator.  The input target selection was based on SDSS $gri$
photometry and has proved to be very efficient with more than 90\% of
the galaxies falling in the desired high redshift range.  More than
two thirds of the galaxies are in the northern slice.  Reliable
redshifts have been obtained for 92\% of the objects observed.  A
large-scale structure of clumps, filaments and voids is apparent in
the most complete regions, similar to what was seen in earlier large
surveys at lower redshift.

Two thirds of the galaxies are in the photometrically well-defined
primary sample and most of the rest (27\%) in the secondary sample.
The primary sample has high spectroscopic completeness in the sense
that good redshifts have been obtained for $\sim90$\% of the targets
within the best-observed fields.  The survey also has high absolute
completeness for LRGs with $z>0.5$ and de-reddened total $i$-band
magnitude brighter than 19.6, since such galaxies lie well clear of
the photometric selection boundaries.  The SDSS-2dF LRG survey thus
offers a significant advance in our understanding of the most massive
galaxies over the past 8~Gyr: the census of such objects must be
almost complete.  It should be possible to distinguish unambiguously
between continuing hierarchical (i.e. merger) and early (high
redshift) models for the formation of LRGs.

For the immediate future, the success of 2SLAQ shows that the new
AAOmega spectrograph for 2dF, which gives a gain of a factor of
up to four in throughput together with better quality and higher
resolution, can be used for cosmological surveys targetting either
higher redshifts or much larger samples, or both.

\subsection*{ACKNOWLEDGEMENTS}

We particularly thank all members of the AAO staff who helped to run
and maintain 2dF, and its complex software, during the course of the
2SLAQ survey.

The 2SLAQ surveys were conducted by a substantial team of people,
coordinated by RCN with ACE and MJD as Principal Investigators for the
LRGs in the UK and Australia respectively.  RDC wishes to acknowledge
their crucial roles, and can claim credit only for contributing to the
observing and data analysis.  He also thanks Will Sutherland and Will
Saunders for helpful discussions about the {\sc Zcode} redshifting
software.

ACE acknowledges support through the Royal Society.  KAP acknowledges
support through an EPSA University of Queensland Research Fellowship
and a UQRSF grant.  RDC is grateful for the hospitality of the
Department of Astrophysics at Oxford, where the final version of this
paper was completed.

Funding for the creation and distribution of the SDSS archive has been
provided by the Alfred P. Sloan Foundation, the Participating
Institutions, NASA, NSF, the US Department of Energy, the Japanese
Monbukagakusho, and the Max Planck Society. The SDSS Web site is
\verb"http://www.sdss.org".

The SDSS is managed by the Astrophysical Research Consortium for the
Participating Institutions. The Participating Institutions are the
University of Chicago, Fermilab, the Institute for Advanced Study, the
Japan Participation Group, The Johns Hopkins University, Los Alamos
National Laboratory, the Max-Planck-Institute for Astronomy, the
Max-Planck-Institute for Astrophysics, New Mexico State University,
the University of Pittsburgh, Princeton University, the United States
Naval Observatory, and the University of Washington.

\appendix

\section{2{\lowercase{d}}F fields and datasets}

Data were obtained for 80 discrete 2dF fields during the 2SLAQ survey,
some with two or more sets of observations.  In general, data obtained
in different observing runs have been treated as independent datasets,
since the target lists were usually modified between runs.  A total of
104 datasets are listed in Table~\ref{fldlist}.

The first seven columns give the name and centre of each field
(equinox J2000).  Column~8 identifies each discrete dataset by field
name and date, usually in the format YYMMXX: these are the datasets
identified in the final redshift catalogues for individual objects
(the `g' distinguishes galaxies from QSOs, while the final `2' denotes
the spectrograph, in practice always no. 2 for the LRGs).  Most fields
were observed on two or more nights during an observing run and all
the data combined, in which case the final two characters are either
`00' or `fi'.  Occasionally data were obtained on only one night, in
which case the last two digits give the date.  A few specially
combined datasets are identified by other character combinations.
Where the same field was observed in different runs the datasets are
listed separately since the target lists will be different, although
often with many objects in common.

Column~9 gives the number of exposures combined in each dataset,
$N_\mathrm{exp}$, normally each of 30~min duration.  Column~10 gives
the mean S/N achieved, averaged over all pixels for all targets, and
$D_\mathrm{mag}$ in column~11 is the mean offset in magnitudes in a
plot of log(counts) versus $r$-band fibre magnitude for each dataset
(total $i_{\rm deV}$ magnitude was used instead of $r_{\rm fib}$ for
some early datasets: these account for most of the negative values of
Dmag).  Both of these are indicators of quality.  The final three
columns give the number of objects in each dataset, the number for
which reliable ($Q>2$) redshifts were obtained and the proportion
these are of the total.  The observations of a field were normally
deemed to be complete when the proportion of reliable redshifts was at
least 85\% (strictly, this criterion was applied to the original set
of targets if some fibres were re-allocated after the first night, but
the numbers quoted here apply to the full datasets).  The footnotes
give further details on some datasets.

\begin{table*}
 \begin{minipage}{250mm}
 \caption{List of 2dF fields observed.}
 \label{fldlist}
 \begin{tabular}{@{}rcccrrrrrrl}
 \hline
 Field & RA &  dec  & dataset   & $N_\mathrm{exp}$ & S/N & $D_\mathrm{mag}$ & obj & $Q>2$ & \% & Notes \\
 \hline
a01 & 08 14 00 & $-$00 12 35 & \verb"a01g_030300_2" & 12 &   6.90 & $-$0.47 & 170 & 159 & 94 & 1,2   \\
a02 & 08 18 00 & $-$00 12 35 & \verb"a02g_033400_2" &  7 &   7.22 & $-$0.94 & 165 & 157 & 95 & 1,2,3 \\
a13 & 09 10 48 & $-$00 12 35 & \verb"a13g_050300_2" &  9 &   3.40 &  0.66 & 200 & 166 & 83 &	      \\
a14 & 09 15 36 & $-$00 12 35 & \verb"a14g_0404fi_2" &  7 &   4.29 &  0.66 & 169 & 159 & 94 &       \\
a15 & 09 20 24 & $-$00 12 35 & \verb"a15g_0404fi_2" &  7 &   5.27 &  0.76 & 178 & 171 & 96 &	      \\
a16 & 09 25 12 & $-$00 12 35 & \verb"a16g_0404fi_2" &  8 &   5.01 &  0.81 & 168 & 147 & 88 &	      \\
a16 & 09 25 12 & $-$00 12 35 & \verb"a16g_050314_2" &  8 &   5.87 &  1.04 & 215 & 209 & 97 &       \\
a17 & 09 30 00 & $-$00 12 35 & \verb"a17g_0403fi_2" &  8 &   5.53 & $-$0.91 & 195 & 186 & 95 &       \\
b01 & 10 01 00 & $-$00 12 35 & \verb"b01g_030400_2" &  8 &   6.63 & $-$0.41 & 142 & 139 & 98 & 2     \\
b00 & 10 02 00 & $-$00 12 35 & \verb"b00g_050417_2" &  4 &   1.67 &  0.15 & 172 & 127 & 74 & 4,5   \\
b02 & 10 05 00 & $-$00 12 35 & \verb"b02g_030300_2" &  8 &   7.25 & $-$0.86 & 163 & 152 & 93 & 1,2   \\
b02 & 10 05 00 & $-$00 12 35 & \verb"b02g_0504fi_2" &  7 &   2.00 &  0.44 & 188 & 136 & 74 & 5     \\
b03 & 10 09 00 & $-$00 12 35 & \verb"b03g_030300_2" & 12 &   6.74 & $-$0.24 & 166 & 152 & 92 & 1,2   \\
b03 & 10 09 00 & $-$00 12 35 & \verb"b03g_0504fi_2" &  8 &   3.91 &  0.84 & 191 & 173 & 91 &       \\
b04 & 10 13 48 & $-$00 12 35 & \verb"b04g_0403fi_2" &  8 &   5.64 & $-$0.90 & 165 & 158 & 96 &       \\
b05 & 10 18 36 & $-$00 12 35 & \verb"b05g_0404fi_2" &  8 &   5.05 &  0.91 & 167 & 153 & 92 &	      \\
b06 & 10 23 24 & $-$00 12 35 & \verb"b06g_0403fi_2" &  8 &   4.70 &  1.00 & 164 & 151 & 92 &	      \\
b06 & 10 23 24 & $-$00 12 35 & \verb"b06g_0504fi_2" &  7 &   4.43 &  0.88 & 219 & 198 & 90 &       \\
b07 & 10 28 12 & $-$00 12 35 & \verb"b07g_0404fi_2" &  9 &   5.56 &  1.23 & 160 & 157 & 98 &	      \\
b08 & 10 33 00 & $-$00 12 35 & \verb"b08g_0404fi_2" & 11 &   4.92 &  1.26 & 171 & 159 & 93 &	      \\
b09 & 10 37 48 & $-$00 12 35 & \verb"b09g_0503fi_2" & 11 &   5.12 &  1.51 & 201 & 184 & 92 &	      \\
b10 & 10 42 36 & $-$00 12 35 & \verb"b10g_0404fi_2" &  8 &   3.81 &  0.84 & 156 & 142 & 91 &	      \\
b11 & 10 47 24 & $-$00 12 35 & \verb"b11g_0503fi_2" &  8 &   4.60 &  1.00 & 206 & 197 & 96 &	      \\
b12 & 10 52 12 & $-$00 12 35 & \verb"b12g_0504fi_2" & 13 &   4.70 &  1.40 & 207 & 190 & 92 &	      \\
b13 & 10 57 00 & $-$00 12 35 & \verb"b13g_050400_2" & 10 &   5.13 &  1.20 & 189 & 166 & 88 &       \\
b14 & 11 01 48 & $-$00 12 35 & \verb"b14g_050467_2" &  8 &   4.66 &  0.85 & 183 & 171 & 93 &	      \\
b15 & 11 06 36 & $-$00 12 35 & \verb"b15g_0504fi_2" & 10 &   3.79 &  0.83 & 187 & 161 & 86 &	      \\
b16 & 11 11 24 & $-$00 12 35 & \verb"b16g_0504fi_2" &  5 &   4.71 &  1.10 & 209 & 193 & 92 &	      \\
c01 & 12 21 30 & $-$00 12 35 & \verb"c01g_030300_2" &  8 &   5.25 & $-$0.62 & 157 & 143 & 91 & 1,2   \\
c00 & 12 22 30 & $-$00 12 35 & \verb"c00g_0505fi_2" &  8 &   4.04 &  0.92 & 196 & 186 & 95 & 4     \\
c02 & 12 25 30 & $-$00 12 35 & \verb"c02g_030300_2" &  8 &   6.87 & $-$0.90 & 168 & 159 & 95 & 1,2   \\
c03 & 12 29 30 & $-$00 12 35 & \verb"c03g_030400_2" &  8 &   5.65 & $-$0.91 & 144 & 130 & 90 & 2     \\
c03 & 12 29 30 & $-$00 12 35 & \verb"c03g_050413_2" &  4 &   4.29 &  0.36 & 168 & 152 & 90 &	      \\
c04 & 12 33 30 & $-$00 12 35 & \verb"c04g_030400_2" &  8 &   6.98 & $-$0.67 & 144 & 141 & 98 & 2     \\
c05 & 12 38 18 & $-$00 12 35 & \verb"c05g_0403fi_2" & 10 &   4.93 & $-$1.00 & 158 & 147 & 93 &       \\
c06 & 12 43 06 & $-$00 12 35 & \verb"c06g_0404rc_2" & 15 &   4.69 &  1.17 & 174 & 151 & 87 &       \\
c07 & 12 47 54 & $-$00 12 35 & \verb"c07g_0403fi_2" &  9 &   4.62 &  0.88 & 173 & 164 & 95 &       \\
d03 & 13 12 00 & $-$00 12 35 & \verb"d03g_040422_2" &  4 &   3.73 &  0.21 & 156 & 142 & 91 &       \\
d04 & 13 16 48 & $-$00 12 35 & \verb"d04g_0503fi_2" & 10 &   4.24 &  0.96 & 203 & 194 & 96 &	      \\
d05 & 13 21 36 & $-$00 12 35 & \verb"d05g_0404fi_2" &  7 &   3.98 &  0.70 & 173 & 149 & 86 &       \\
d06 & 13 26 24 & $-$00 12 35 & \verb"d06g_0403fi_2" & 11 &   5.12 & $-$0.79 & 169 & 163 & 96 &       \\
d07 & 13 31 12 & $-$00 12 35 & \verb"d07g_0404fi_2" &  7 &   3.34 &  0.41 & 161 & 113 & 70 &       \\
d07 & 13 31 12 & $-$00 12 35 & \verb"d07g_0504fi_2" &  9 &   5.15 &  1.15 & 207 & 192 & 93 &       \\
d08 & 13 36 00 & $-$00 12 35 & \verb"d08g_030406_2" &  5 &   6.68 & $-$1.17 & 165 & 165 &100 & 2     \\
d08 & 13 36 00 & $-$00 12 35 & \verb"d08g_0505fi_2" &  8 &   5.28 &  1.05 & 205 & 201 & 98 &       \\
d09 & 13 40 00 & $-$00 12 35 & \verb"d09g_030300_2" & 16 &   5.60 & $-$0.22 & 165 & 153 & 93 & 1,2,6 \\
d10 & 13 40 00 & $-$00 12 35 & \verb"d10g_030400_2" & 11 &   6.03 & $-$0.31 & 166 & 163 & 98 & 2,6   \\
d10 & 13 40 00 & $-$00 12 35 & \verb"d10g_0504fi_2" &  9 &   3.83 &  0.76 & 193 & 170 & 88 & 6     \\
d11 & 13 44 48 & $-$00 12 35 & \verb"d11g_050467_2" &  9 &   4.31 &  0.95 & 186 & 168 & 90 &	      \\
d12 & 13 49 36 & $-$00 12 35 & \verb"d12g_0504fi_2" & 10 &   3.76 &  0.84 & 231 & 172 & 74 & 7     \\
d13 & 13 54 24 & $-$00 12 35 & \verb"d13g_0503fi_2" &  8 &   7.27 &  1.08 & 190 & 175 & 92 &	      \\
d14 & 13 59 12 & $-$00 12 35 & \verb"d14g_0404fi_2" &  6 &   3.09 &  0.55 & 155 & 107 & 69 &       \\
d14 & 13 59 12 & $-$00 12 35 & \verb"d14g_0504fi_2" & 10 &   3.91 &  0.92 & 201 & 182 & 91 &       \\
d15 & 14 04 00 & $-$00 12 35 & \verb"d15g_0503fi_2" &  7 &   4.26 &  1.02 & 197 & 176 & 89 &	      \\
d16 & 14 08 48 & $-$00 12 35 & \verb"d16g_0404rc_2" & 12 &   4.80 &  1.41 & 176 & 160 & 91 &       \\
d17 & 14 13 36 & $-$00 12 35 & \verb"d17g_0503fi_2" & 12 &   4.93 &  1.29 & 180 & 172 & 96 &	      \\
			       
e01 & 14 34 00 & $-$00 12 35 & \verb"e01g_030400_2" &  8 &   4.37 & $-$1.18 & 147 & 138 & 94 & 2     \\
e01 & 14 34 00 & $-$00 12 35 & \verb"e01g_050411_2" &  4 &   3.99 &  0.37 & 165 & 155 & 94 &	      \\
e02 & 14 38 00 & $-$00 12 35 & \verb"e02g_030406_2" &  4 &   4.65 & $-$1.55 & 156 & 130 & 83 & 2     \\
e02 & 14 38 00 & $-$00 12 35 & \verb"e02g_050413_2" &  5 &   3.51 &  0.45 & 143 & 134 & 94 &       \\
e03 & 14 42 48 & $-$00 12 35 & \verb"e03g_050400_2" & 15 &   4.12 &  1.28 & 197 & 182 & 92 &	      \\
e04 & 14 47 36 & $-$00 12 35 & \verb"e04g_0404fi_2" & 11 &   3.81 &  0.94 & 166 & 145 & 87 &	      \\
e04 & 14 47 36 & $-$00 12 35 & \verb"e04g_0505fi_2" &  7 &   4.33 &  0.83 & 214 & 194 & 91 &       \\
e05 & 14 52 24 & $-$00 12 35 & \verb"e05g_0504fi_2" & 10 &   4.25 &  1.11 & 184 & 165 & 90 &	      \\
 \end{tabular}
 \end{minipage}
\end{table*}
\begin{table*}
 \begin{minipage}{250mm}
 \contcaption{}
 \begin{tabular}{@{}rcccrrrrrrl}
 \hline
 Field & RA &  dec  & dataset   & $N_\mathrm{exp}$ & S/N & $D_\mathrm{mag}$ & obj & $Q>2$ & \% & Notes \\
 \hline
e06 & 14 57 12 & $-$00 12 35 & \verb"e06g_0404fi_2" &  9 &   3.94 &  0.93 & 154 & 121 & 79 &       \\
e06 & 14 57 12 & $-$00 12 35 & \verb"e06g_0504fi_2" &  9 &   3.72 &  0.91 & 196 & 167 & 85 &       \\
e07 & 15 02 00 & $-$00 12 35 & \verb"e07g_050508_2" &  4 &   3.21 &  0.17 & 169 & 129 & 76 &       \\
e07 & 15 02 00 & $-$00 12 35 & \verb"e07g_050730_2" &  8 &   3.97 &  0.94 & 231 & 209 & 90 &       \\
e08 & 15 06 48 & $-$00 12 35 & \verb"e08g_050417_2" &  5 &   2.73 &  0.16 & 170 & 136 & 80 & 7     \\
e08 & 15 06 48 & $-$00 12 35 & \verb"e08g_050802_2" & 10 &   4.03 &  1.02 & 219 & 184 & 84 &       \\
e09 & 15 11 36 & $-$00 12 35 & \verb"e09g_050730_2" &  6 &   3.39 &  0.49 & 198 & 150 & 76 &       \\
e10 & 15 16 24 & $-$00 12 35 & \verb"e10g_050801_2" &  7 &   3.62 &  0.63 & 171 & 145 & 85 &       \\
    &       	 &   &   &    &     	       &    &	     &	     &	   &	 	      \\
s01 & 20 57 36 & $-$00 15 00 & \verb"s01g_030800_2" &  4 &   2.98 & $-$0.47 & 170 &  95 & 56 & 8,9   \\
			       
s06 & 21 21 36 & $-$00 15 00 & \verb"s06g_030900_2" & 12 &   5.53 &  0.58 & 171 & 155 & 91 &	      \\
s07 & 21 26 24 & $-$00 15 00 & \verb"s07g_0410fc_2" & 10 &   3.43 &  0.53 & 187 & 159 & 85 &       \\
s07 & 21 26 24 & $-$00 15 00 & \verb"s07g_050730_2" &  9 &   3.83 &  0.69 & 206 & 169 & 82 &       \\
s08 & 21 31 12 & $-$00 15 00 & \verb"s08g_04se69_2" & 18 &   4.07 & $-$1.10 & 187 & 148 & 79 &       \\
s09 & 21 36 00 & $-$00 15 00 & \verb"s09g_041013_2" &  3 &   2.20 & $-$0.54 & 152 &  81 & 53 &       \\
s09 & 21 36 00 & $-$00 15 00 & \verb"s09g_050801_2" &  8 &   4.38 &  0.84 & 166 & 156 & 94 &       \\
s10 & 21 40 48 & $-$00 15 00 & \verb"s10g_050508_2" &  3 &   2.25 & $-$0.20 & 172 & 111 & 65 &       \\
s10 & 21 40 48 & $-$00 15 00 & \verb"s10g_050729_2" &  7 &   2.86 &  0.51 & 187 & 139 & 74 &       \\
s11 & 21 45 36 & $-$00 15 00 & \verb"s11g_050731_2" &  9 &   3.61 &  0.94 & 204 & 167 & 82 &       \\
s12 & 21 50 24 & $-$00 15 00 & \verb"s12g_030800_2" & 13 &   4.94 &  0.11 & 168 & 152 & 90 & 8     \\
			       
s25 & 22 52 48 & $-$00 15 00 & \verb"s25g_030800_2" & 14 &   6.24 &  0.73 & 169 & 163 & 96 & 8     \\
s25 & 22 52 48 & $-$00 15 00 & \verb"s25g_030920_2" &  4 &   2.86 & $-$0.93 & 166 & 135 & 81 &       \\
s26 & 22 57 36 & $-$00 15 00 & \verb"s26g_041013_2" &  4 &   2.80 &  0.04 & 148 & 119 & 80 &       \\
s26 & 22 57 36 & $-$00 15 00 & \verb"s26g_050801_2" &  4 &   3.25 &  0.14 & 164 & 130 & 79 &       \\
s27 & 23 02 24 & $-$00 15 00 & \verb"s27g_030900_2" &  8 &   3.79 & $-$0.09 & 166 & 148 & 89 &       \\
s28 & 23 07 12 & $-$00 15 00 & \verb"s28g_04oc89_2" & 11 &   3.17 &  0.71 & 147 & 122 & 83 &       \\
s29 & 23 12 00 & $-$00 15 00 & \verb"s29g_04com1_2" & 22 &   4.17 & $-$0.61 & 169 & 147 & 87 &       \\
s30 & 23 16 48 & $-$00 15 00 & \verb"s30g_0410fi_2" & 10 &   3.45 &  0.64 & 170 & 156 & 92 &       \\
s31 & 23 21 36 & $-$00 15 00 & \verb"s31g_050730_2" &  8 &   3.97 &  0.92 & 187 & 169 & 90 &       \\
s32 & 23 26 24 & $-$00 15 00 & \verb"s32g_050801_2" &  8 &   3.99 &  0.76 & 187 & 161 & 86 &       \\
			       
s47 & 00 38 24 & $-$00 15 00 & \verb"s47g_050801_2" &  4 &   2.88 &  0.08 & 187 & 150 & 80 &       \\
s48 & 00 43 12 & $-$00 15 00 & \verb"s48g_046nts_2" & 22 &   3.24 & $-$1.06 & 204 & 182 & 89 &       \\
s49 & 00 48 00 & $-$00 15 00 & \verb"s49g_041013_2" &  4 &   2.69 & $-$0.24 & 158 & 134 & 85 &       \\
s50 & 00 52 48 & $-$00 15 00 & \verb"s50g_030800_2" & 12 &   5.93 &  0.58 & 167 & 158 & 91 & 8     \\
s50 & 00 52 48 & $-$00 15 00 & \verb"s50g_030920_2" &  4 &   2.91 & $-$0.81 & 172 & 143 & 83 &       \\
s51 & 00 57 36 & $-$00 15 00 & \verb"s51g_0410fi_2" & 12 &   3.52 &  0.64 & 190 & 168 & 88 &       \\
s52 & 01 02 24 & $-$00 15 00 & \verb"s52g_030900_2" & 12 &   4.43 &  0.21 & 172 & 155 & 90 &       \\
s67 & 02 14 24 & $-$00 15 00 & \verb"s67g_040000_2" & 23 &   3.95 & $-$0.73 & 162 & 140 & 86 &       \\
s68 & 02 19 12 & $-$00 15 00 & \verb"s68g_041013_2" &  4 &   2.36 &  0.04 & 148 & 121 & 82 &       \\
s69 & 02 24 00 & $-$00 15 00 & \verb"s69g_030900_2" & 11 &   4.44 &  0.21 & 165 & 141 & 85 &       \\
s70 & 02 28 48 & $-$00 15 00 & \verb"s70g_0410fi_2" & 13 &   3.44 &  0.77 & 172 & 149 & 87 &       \\
\\			      
 \end{tabular}		     
			       
Notes to Table~A1	       
\vspace{3mm}

1. A central wavelength of 6350\AA\ was used in March 2003 for fields
a01, a02, b02, b03, c01, c02 and d09.  All subsequent data used\\
\hspace*{3mm} 6150\AA.

2. The target selection in March and April 2003 had an $i_{\rm deV}$
magnitude limit of 19.5.  This applies to the fields listed in note~1 plus\\ 
\hspace*{3mm} b01, c03, c04, d08, e01 and e02.  Most of these fields were
re-observed subsequently with the standard $i_{\rm deV} = 19.8$ limit.  However,\\ 
\hspace*{3mm} a01 and a02 in particular should be omitted from statistical
analyses since they have low primary Sample~8 completeness.

3. The redshifts for a02 come from the combined frame for both months,
but for some purposes the separate spectra from March or \\
\hspace*{3mm} April 2003 may be better.

4. b00 and c00 are hybrid field centres, shifted 15~arcmin (1~min in
RA) east from b01 and c01 respectively, to increase overlap with\\ 
\hspace*{3mm} b02 and c02.

5. The April 2005 data for b00 and b02 were badly affected by twilight
and moonlight; data from the last night only were used for\\ 
\hspace*{3mm} most b00 objects.

6. d09 and d10 have the same field centre but a different selection of
targets, with many faint objects ($19.5 < i_{\rm deV} < 20$) in d10.

7. There was bad scattered light halation for d12 and e08 in April
2005; the overall $z$ yield for d12 was lowered by many weak spectra\\
\hspace*{3mm} observed on the final night only.

8. The secondary Sample~9 targets were inadvertently given higher
priority than the primary Sample~8 targets for fields s01, s12, s25\\
\hspace*{3mm} and s50 in August 2003.  s25 and s50 were re-observed in September
2003 but s01 and s12 have very low Sample~8 completeness.

9. s01 has very low redshift completeness and should be omitted from
statistical analyses.  This field is at low Galactic latitude and\\
\hspace*{3mm} suffers from significant interstellar extinction and high 
($\sim~20$\%) stellar contamination.

 \end{minipage}
\end{table*}

\section{The 2SLAQ redshifting procedure}

This appendix gives more information specific to the redshifting of the
2SLAQ LRGs, in particular regarding the template spectra.

\subsection{The template spectra}

\begin{figure}
 \includegraphics[width=85mm]{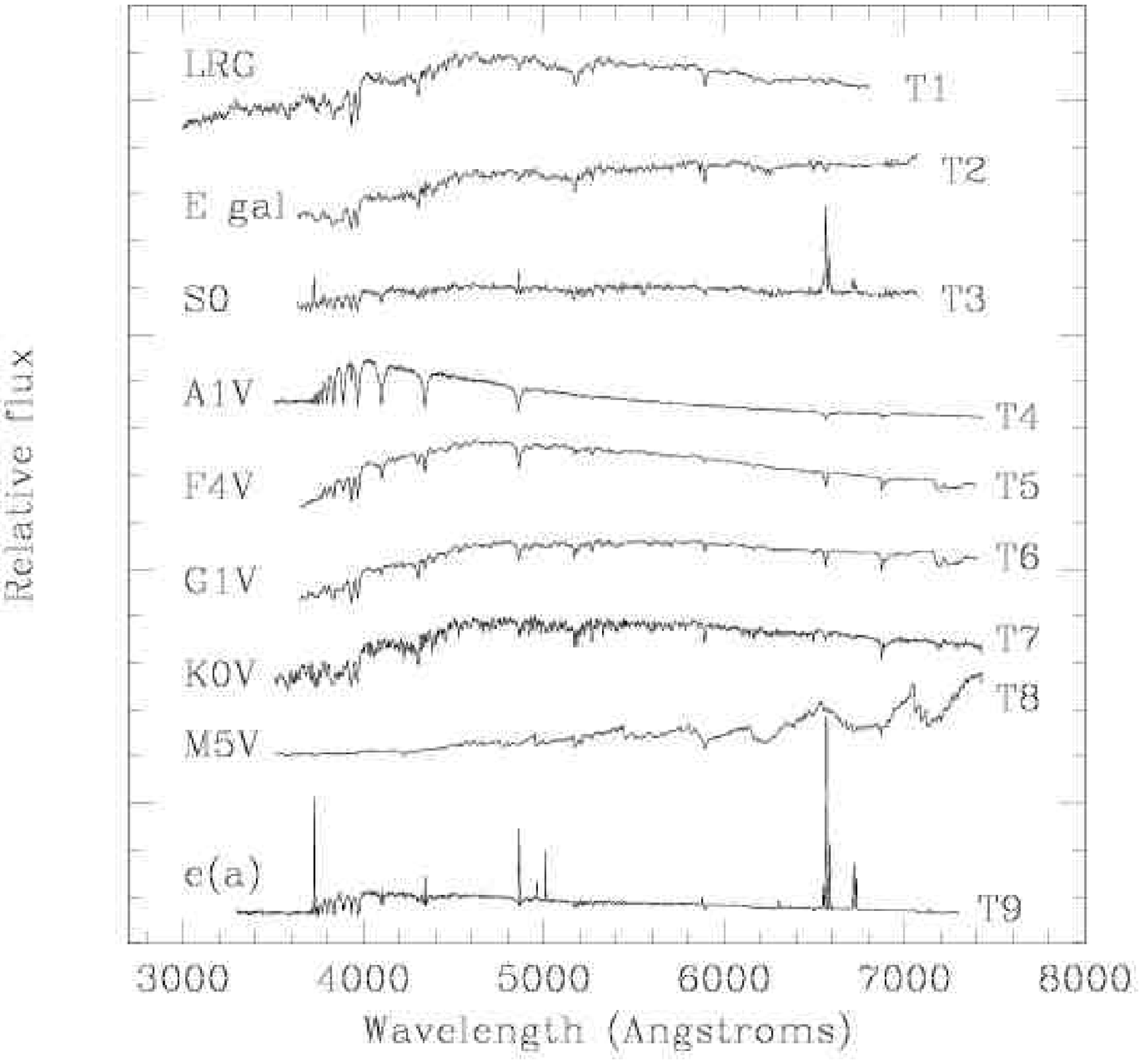}
 \caption{The nine redshift template spectra T1 to T9 from top to
  bottom, plotted with arbitrary flux scales and offsets.  The first
  and last are composite spectra, the second and third are nearby
  galaxies, and the remaining five are a sequence of stars from type A
  to M.}
 \label{tempspec}
\end{figure}

Nine template spectra were used to derive redshifts for the LRGs, as
listed in Table~\ref{templist} and plotted in Fig.~\ref{tempspec}.  T1
is a composite LRG spectrum derived from the somewhat lower redshift
LRGs found in the SDSS LRG survey \citep{eis03}: this gives the best
fit to at least 65\% of the 2SLAQ sample and a reliable redshift for
more than 90\%, since it is both well-matched to the data and extends
down to 3000\AA\ rest wavelength.  The second and third are bright
galaxies used for the 2dFGRS, NGC~3379 which is a standard E1
elliptical galaxy and NGC~5248, an S0 with emission lines.  The next
five templates are representative stars covering the spectral classes
A to M, three of them in common with the 2dFGRS.  T9 is a composite
e(a) spectrum (from Paul Hewett).

\begin{table}
 \caption{The template spectra.}
 \label{templist}
 \begin{tabular}{@{}clrcr}
  \hline
Template & Name &  Velocity  & $z$  & \%age \\
  & & (km\,s$^{-1}$) & \\
  \hline
T1 & lrg\_te\_s &    85 &  0.50 & 60.6 \\
T2 & NGC3379b   &  $-$115 &  0.50 & 16.2 \\
T3 & NGC5248b   &   $-$90 &  0.50 &  8.2 \\
T4 & a1star\_b  &    60 &  0.00 &  0.3 \\
T5 & E323\_F4V  &    85 &  0.00 &  2.8 \\
T6 & E329\_G1V  &    10 &  0.00 &  3.0 \\
T7 & k0star\_b  &  $-$130 &  0.00 &  0.8 \\
T8 & m5star\_b  &  $-$115 &  0.00 &  0.0 \\
T9 & pch\_t1\_s &   155 &  0.50 &  8.1 \\
  \hline
 \end{tabular}
\end{table}

The third column of Table~\ref{templist} lists small zero-point
velocity offsets which have been derived in two ways: from the mean
differences for a set of LRGs, all cross-correlated with each
template, and by cross-correlating each template with a synthetic
K-type stellar spectrum generated using the {\it ssg} code of
\citet{bg78}.  The two techniques gave consistent results to within
$\sim20$\,km\,s$^{-1}$, a small error compared with the accuracy of
$\sim100$\,km\,s$^{-1}$ for individual galaxy redshifts.  These
relative velocities are all tied to the composite LRG template T1,
whose 85\,km\,s$^{-1}$ velocity is the difference between the vacuum
wavelengths used by the SDSS and the air wavelengths used by {\sc
2dfdr}.  The effects on the final redshifts are of course small,
amounting to no more than 0.0003.

The fourth column of Table~\ref{templist} is a redshift offset applied
to some templates, to optimise the interactive display for most of the
targets.  The four galaxies were shifted to $z = 0.5$, the mean
redshift of the LRGs, while the stars were left at $z = 0$.  The
derived redshifts were not affected significantly by these offsets.
In practice the only stars which contaminate the LRG sample are M-type
dwarfs (5\% of the targets), but the G and K-type stellar spectra
often provide useful confirmation of the galaxy redshifts, especially
for $z<0.4$.  The final column in Table~\ref{templist} gives the
percentage of galaxies for which each template was used.

\subsection{Usage of templates}

\begin{figure}
 \includegraphics[width=85mm]{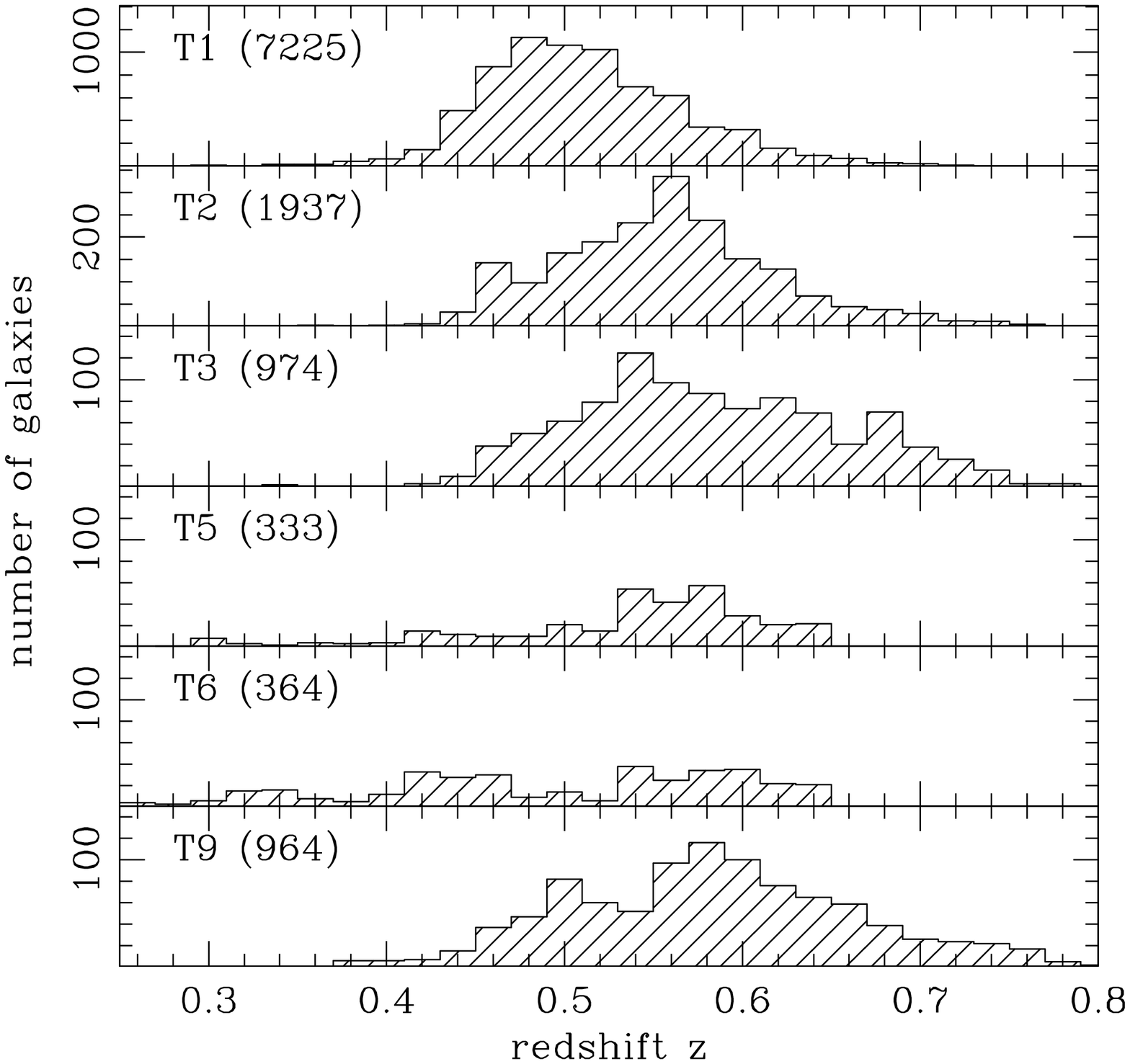}
 
 \caption{Histograms of the redshifts for which each of the templates
 was used, for Sample~8 and 9 galaxies, omitting the little-used A-
 and K-type spectra and the M-type spectrum which fitted only
 foreground M-stars.  The numbers of galaxies in each panel are given
 in parentheses: note the different ordinate scales in the top two
 panels.}

 \label{tzhist}
\end{figure}

The template which gives the best redshift estimate for each galaxy is
a function of galaxy type.  Fig.~\ref{tzhist} shows the redshift
distribution for most of the templates.  Three of the templates, T4,
T7 and T8 for A-, K- and M-type stars respectively, have been omitted
because they rarely if ever gave the best fit to a galaxy.  Not
surprisingly, the mean SDSS LRG spectrum (T1) provides the best fit to
by far the largest fraction of the targets, about 60\% of the total.
The NGC~3379 (T2) spectrum, which is very similar but without the UV
tail, was used for a further 16\%: for most objects these two
templates give virtually identical results.  Thus at least three
quarters of the 2SLAQ sample matches very well the spectrum of a lower
redshift quiescent elliptical galaxy or LRG.  Note that the ordinate
scales in Fig.~\ref{tzhist} are different for these two templates.

The other two galaxy templates, the emission-line S0 galaxy NGC~5248
(T3) and a `k+a+em' composite (T9), are each used for 8\% of the
sample.  These include a higher proportion of active galaxies, either
with active galactic nuclei (AGNs) or star-forming (SF) (these two
classes cannot be easily distinguished in the 2SLAQ spectra, since
often the only clue is the presence of [O~II] at 3727\AA\/), and they
tend to be at higher redshifts.  Note that the emission lines in T3
and T9 are blanked out for the cross-correlations and do not
contribute to the redshift determinations.  For the great majority of
galaxies with reliable redshifts, all four galaxy templates agree.

The F-, G- and K-type stellar templates together give the best fit to
about 6\% of the LRGs, although again the galaxy templates usually
give the same redshifts.  There is a tendency for the stars to be used
for the lowest redshift galaxies with $0.3<z<0.4$, perhaps because
such galaxies are not true LRGs.  The upper cut-off at $z=0.65$ for T5
and T6 is an artefact of the {\sc Zcode}.  The A-type stellar spectrum
gives an apparent best fit in only $\sim0.2$\% of cases and is often
clearly wrong when the fit is checked visually.  The M-type spectrum
is used only for foreground M-dwarfs.

\label{lastpage}

\end{document}